\documentclass[a4paper,onecolumn,allowtoday,unpublished]{quantumarticle}
\pdfoutput=1
\usepackage[T1]{fontenc}
\usepackage[british]{babel}
\usepackage[unicode,colorlinks]{hyperref}
\usepackage{graphicx}
\usepackage{tabularx}
\usepackage{amsmath,amssymb,amsthm,bm}
\usepackage{xspace}
\usepackage{mathtools}
\usepackage{dsfont}
\usepackage[dvipsnames]{xcolor}
\usepackage{stmaryrd}
\usepackage{tikz,pgfplots}\pgfplotsset{compat=1.18}

\pgfplotsset{
    legend image with text/.style={
        legend image code/.code={%
            \node[anchor=center] at (0.3cm,0cm) {#1};
        }
    },
}

\usepackage{subfig}
\usepackage{lipsum}
\usepackage{mathrsfs}
\usepackage[all]{xy}

\usepackage{csquotes}
\usepackage[backend=bibtex,style=phys,biblabel=brackets,eprint=true,doi=false,url=true,maxnames=10,giveninits=true]{biblatex}
\DeclareMathOperator{\cl}{cl}

\let\originalleft\left
\let\originalright\right
\renewcommand{\left}{\mathopen{}\mathclose\bgroup\originalleft}
\renewcommand{\right}{\aftergroup\egroup\originalright}

\newcommand{\g}{\mc G}
\newcommand{\z}{\mc Z}
\newcommand{\zg}{\mc Z \circ \mc G}
\newcommand{\gmap}{\widehat{\mc G}}

\newcommand{\zgmap}{\widehat{\mc{ZG}}}
\newcommand{\pkey}{p^\mathrm{K}}

\newcommand{\mkey}[0]{\mathcal{M}^{\mathrm{K}}}

\newcommand{\mtest}[0]{\mathcal{M}^{\mathrm{T}}}

\newcommand{\bra}[1]{\left\langle #1 \right|}
\newcommand{\ket}[1]{\left| #1 \right\rangle}
\newcommand{\braket}[2]{\left\langle #1 \middle| #2 \right\rangle}
\newcommand{\ketbra}[2]{\left|#1\middle\rangle\middle\langle#2\right|}

\newcommand{\De}[1]{\left[#1\right]}

\newtheorem{theorem}{Theorem}[section]
\newtheorem*{theorem*}{Theorem}

\newtheorem{proposition}[theorem]{Proposition}

\newtheorem{remark}{Remark}

\def\Tr{\operatorname{Tr}}

\usepackage{booktabs}

\newcommand{\pro}[1]{\ket{#1}\bra{#1}}
\newcommand{\mc}[1]{\mathcal{#1}}

\hypersetup{
  pdftitle={Variable-length MDI QKD under source imperfections and local randomization over horizontal links}
}
\title{Variable-length MDI QKD under source imperfections and local randomization over horizontal links}
\begin{document}

\author[1,2]{Gabriele Staffieri\textsuperscript{*}}
\email{g.staffieri2@phd.uniba.it}
\orcid{0009-0004-8312-593X}

\author[3]{Mart\'i Sales-Moragues\textsuperscript{*}}
\email{marti.sales@luxquanta.com}

\author[3]{Carlos Pascual-Garc\'ia}
\orcid{0000-0003-0659-7349}

\affil[3]{Luxquanta Technologies S.L., Av. Joan Carles I, 30, 1º1ª, 08908 L'Hospitalet de Llobregat, Barcelona, Spain}

\affil[1]{Dipartimento Interateneo di Fisica, Università degli Studi di Bari Aldo Moro, 70126 Bari, Italy}

\affil[2]{Istituto Nazionale di Fisica Nucleare (INFN), Sezione di Bari, 70126 Bari, Italy}

\maketitle

\begingroup
\renewcommand{\thefootnote}{\fnsymbol{footnote}}
\footnotetext[1]{These authors contributed equally to this work.}
\endgroup

\begin{abstract}
Quantum key distribution (QKD) provides security guarantees that hold despite device imperfections and finite-size effects. We address these challenges in measurement-device-independent (MDI) QKD, combining its intrinsic untrusted detection with an imperfection model for Alice's and Bob's sources via fidelity bounds. We establish variable-length general security under the Marginal-constrained Entropy Accumulation Theorem (MEAT), estimating the relevant Rényi entropies through conic optimization. We evaluate our model under a horizontal channel link, reporting secret keys for blocks as small as $n\sim3\times10^4$ rounds under atmosphere-free conditions, and show that compensating for unequal channel attenuation through appropriately chosen transmitted amplitudes improves performance in asymmetric configurations, enabling communication distances beyond $80$ km for commercial satellite systems.
\end{abstract}

\section{Introduction}

Quantum Key Distribution (QKD) \cite{BB84,E91,Renner2005} enables two honest parties, traditionally called Alice and Bob, to establish a secret key using an insecure quantum channel that can be eavesdropped by an adversary, commonly refereed to as Eve. QKD provides information-theoretic security for the communications of Alice and Bob, i.e. a security thanks the principles of quantum mechanics  even against the most powerful tampering that Eve can perform, which are the coherent attacks \cite{Renner2022Security,ferradini2025definingsecurityquantumkey,tupkary2025qkdsecurityproofsdecoystate}. Therefore, proving security against these type of attacks is considered the greatest accomplishment of any QKD security proof.

Within this context, the Marginal-constrained Entropy Accumulation Theorem (MEAT) \cite{arqand2025MEAT} represents one the most advanced framework for the theoretical analysis of QKD, providing general security for all device-dependent QKD implementations. The MEAT generalizes previous results in entropy accumulation \cite{dupuis2016entropy,dupuis2019entropy,metger2022generalised,arqand2024generalized}, and allows a distillation of secret keys under real-time operation \cite{bodenstedt2026}, while enhancing the performance of QKD via an adaptive, variable-length approach \cite{Tupkary2024,tupkary2025qkdsecurityproofsdecoystate}.

Building on the MEAT framework, a finite-size security analysis for discrete variable protocols \cite{Pirandola_2020} like decoy-state BB84 was performed \cite{kamin25MEATsecurity,tupkary2026rigorouscompletesecurityproof}, while subsequent works \cite{Navarro_finite,Navarro2025,pascualgarcia2026RPSK} incorporated the MEAT into continuous variable QKD protocols \cite{zhang2024continuousvariable}, together with new, conic optimization schemes introduced in \cite{Lorente2025quantumkey}. Despite these advances, the application of the MEAT to measurement-device-independent (MDI) QKD protocols remains unexplored. MDI QKD enables Alice and Bob to establish a secret key through an intermediate untrusted relay, whose measurements can be entirely controlled by Eve without compromising the protocol security \cite{LoCurtyQi2012}. This removes the vulnerabilities associated with attacks exploiting imperfections in the measurement devices\cite{lydersen2010hacking,tanner2014optimised}. Previous studies have established finite-key security for fiber-based MDI QKD protocols, including security against general coherent attacks based on entropic uncertainty relations \cite{Curty2014FiniteKey} or Gaussian de Finetti reductions \cite{Lupo2018CoherentMDI}. Other finite-size analyses have considered collective Gaussian attacks \cite{Papanastasiou2017FiniteSize}, while free-space MDI QKD has been investigated both experimentally \cite{Cao2020FreeSpaceMDI} and through collective security analyses \cite{Ghalaii2023FreeSpaceMDI,Yao2025FluctuatingMDI}.

In this work, we consider horizontal free-space communication links at different altitudes, motivated by potential applications involving ground-based terminals, drones, high-altitude platforms \cite{Karakosta2025}, and satellite systems \cite{Sidhu2021,Chiti2024}. We characterize atmospheric extinction using LOWTRAN \cite{Kneizys1988LOWTRAN,PicciaLowtran} and turbulence through the Hufnagel--Valley model \cite{Hufnagel1974,Valley1980,AndrewsLaserBeam}, while accounting for diffraction, beam wandering, and pointing fluctuations in the channel transmittance \cite{recolons2007beamwander,vasylyev2012toward,pirandola2021limits}. Our analysis is focused on the weak-turbulence regime for atmospheric propagation and also considers the atmosphere-free limit, relevant to outer space optical links. 

Furthermore, to account for realistic implementations, we incorporate source imperfections into our security analysis. Specifically, we adopt a partial source-characterization model \cite{Curras2025numerical,pereira2025optimalkeyratesquantum,wiesemann2026}, in which fidelity constraints bound the deviations of the actual emitted states from their ideal reference states. This is motivated by the fact that, although MDI QKD removes detector characterizations, real sources suffer from state-preparation flaws and side channels that an eavesdropper could exploit. Thus, bounding the deviations through fidelity constraints closes this gap, while handling the fact that a full characterization of the source is also unrealistic.

Finally, we investigate the effect of local randomization, a trusted preprocessing procedure in which private noise is introduced into the raw key to potentially improve its security \cite{Kraus2005,Renner2005}. Following the treatment developed in \cite{staffieri2026}, we incorporate local randomization into our MEAT-based security analysis and evaluate its impact on the achievable secret-key rates.

The rest of this manuscript goes as follows. Section \ref{sec:Prolegomena} provides the necessary mathematical tools, Section \ref{sec:ProtDescription} describes our MDI protocol, which we characterize according to the MEAT and an imperfect model for the source in Section \ref{sec:FiniteSize}. Section \ref{sec:Horizontal} provides the description for the horizontal quantum link, which yields diverse results that we describe in Sections \ref{Sec:numerics} and \ref{Sec:Results}, and discuss in \ref{Sec:Discussion}.

\section{Mathematical prolegomena}\label{sec:Prolegomena}

Let  $\mc{H}_X$ a Hilbert space for a register $X$, and $X_1^n$ a collection of $n$ copies with the same embedding on $X$ and $\mc{D}(X)$ the set of all density matrices defined for such register. For a state $\rho_{XY}$, we denominate it as classical-quantum (cq-state) whenever it admits the description
\begin{align}
    \rho_{XY} &= \sum_{x\in \mathcal{X}} p(x) \ketbra{x}{x}_X\otimes \rho^x_Y \\
    &= \sum_{x\in \mathcal{X}} \ketbra{x}{x}_X\otimes \rho_{Yx}, 
\end{align}
on some countable alphabet $\mathcal{X}$, which defines a distribution on a probability space $\mathbb{P}_{\mc{X}}$. We define a probabilistic event as the subset $\Omega \subseteq \mathcal{X}$. This event induces a conditioning for cq-states
\begin{equation} \label{eq:TrPr}
    \rho_{XY|\Omega} = \frac{1}{\mathrm{Pr}_\rho [\Omega]}\sum_{x\in \Omega} p(x) \ketbra{x}{x}_X\otimes \rho^x_Y,
\end{equation}
with $\mathrm{Pr}_\rho[{\Omega}] = \sum_{x\in {\Omega}} p(x)$ representing the probability of observing the event. Regarding entropic quantities, we will focus our analysis on the $f-$weighted Rényi entropy of order $\alpha$ \cite{arqand2024generalized,arqand2025MEAT}. Given the ccq-state $\rho \in \mc{D}(SCE)$ with $\rho_c = p(c) \in \mathbb{P}_\mathcal{C}$ we define
\begin{align}
     H^{\uparrow,f}_\alpha (S|{C}E)_{\rho} = \frac{\alpha}{1-\alpha} \log \left( \sum_{{c}\in \mc{C}} p({c}) 2^{\frac{1-\alpha}{\alpha} (-f ({c})+ H^\uparrow_\alpha (S|E))_{\rho_{|{c}}}}  \right).
\end{align}
where $\alpha \in \mathbb{R}_> \backslash\{1\}$, and we use the conditional sandwiched Rényi entropy \cite{tomamichel2015quantum}

\begin{equation}
    H^\uparrow_{\alpha}(S|E)_\rho = \sup_{\sigma_E \in \mc{D}(E)} - D_\alpha(\rho_{SE}\|\mathds{1}_S\otimes\sigma_E),
\end{equation}
with the sandwiched Rényi divergence
\begin{equation} \label{eq:QRenyiDiv}
    D_\alpha(\rho\|\sigma) = \frac{1}{\alpha-1}\log\De{\frac{\Tr\left[(\sigma^\frac{1-\alpha}{2\alpha}\rho\sigma^\frac{1-\alpha}{2\alpha})^\alpha\right]}{\Tr[\rho]}},  
\end{equation}
whenever $\alpha \in \mathbb{R}_> \backslash\{1\}$ and $\mathrm{supp}(\rho) \subseteq \mathrm{supp}(\sigma)$. For the sake of simplifying the notation, we will commonly use

\begin{align}\label{eq:PsiFunction}
    \Psi_\alpha(\rho,\sigma) := \Tr\left[(\sigma^\frac{1-\alpha}{2\alpha}\rho\sigma^\frac{1-\alpha}{2\alpha})^\alpha\right],
\end{align}
which is the quasi-relative Rényi entropy. 
We also introduce the down-arrow conditional sandwiched Rényi entropy,
defined as
\begin{equation}
    H^\downarrow_{\alpha}(S|E)_\rho
    =
    -D_\alpha\left(
        \rho_{SE}\middle\|
        \mathds{1}_S\otimes\rho_E
    \right),
\end{equation}
where $\rho_E=\Tr_S[\rho_{SE}]$.
Unlike the up-arrow entropy, the down-arrow entropy does not
involve an optimization over the conditioning state.
We will also make use of the Kullback-Leibler divergence $D_\mathrm{KL}(\cdot \| \cdot)$, which is a particular instance  of the Rényi relative entropy when $\alpha \to 1$ and both states are classical.

\section{Protocol description}\label{sec:ProtDescription}

\begin{figure}[!h]
    \centering
    \includegraphics[width=1.0\linewidth]{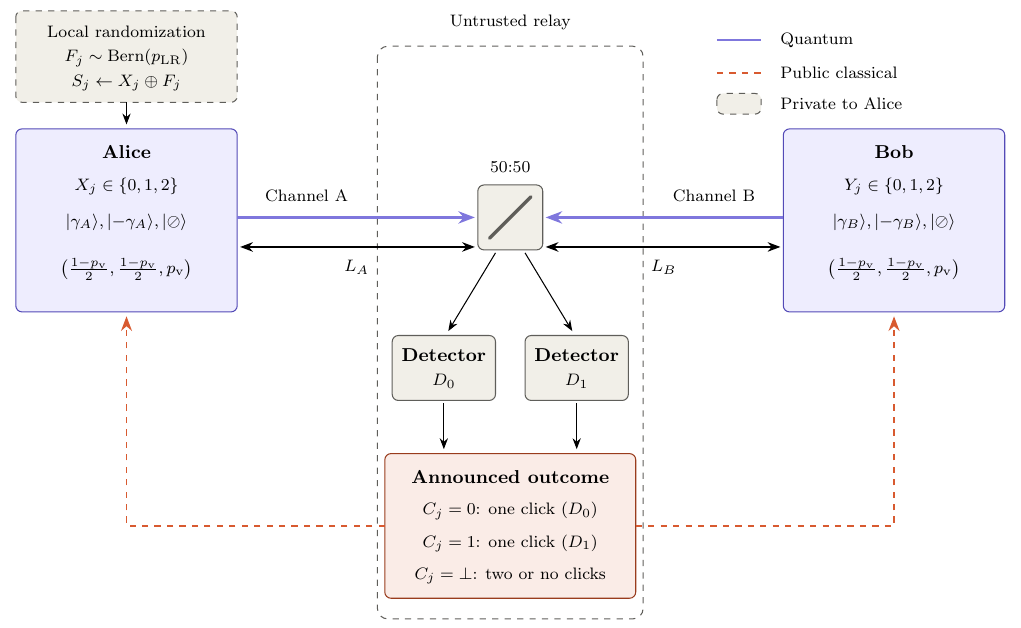}
    \caption{Scheme of the MDI QKD protocol. Alice and Bob prepare coherent states with amplitudes $\gamma_A$ and $\gamma_B$ and transmit them through channels of lengths $L_A$ and $L_B$ to an untrusted relay, which announces the detection outcome $C_j\in\{0,1,\bot\}$. Based on the relay outcome, Alice and Bob select the rounds for key generation or parameter estimation, with eligible rounds retained for key generation with probability $p_{\mathrm K}$. Following parameter estimation, Alice applies local randomization to her raw-key register $S_j$, flipping each retained bit with probability $p_{\mathrm{LR}}$. The protocol concludes with information reconciliation and privacy amplification.}
    \label{fig:protocol}
\end{figure}

Let us start with a description of our MDI QKD protocol, which is amenable to a finite-size analysis under the MEAT \cite{arqand2025MEAT} and graphically described in Figure \ref{fig:protocol}. Remarkably, we explicitly note that the preparations by Alice and Bob are considered to be imperfect, and only roughly correspond to their expected inputs. We rigorously address this difference in Subsection \ref{Subsec:sourceImperfections}.

\begin{itemize}
    \item \textit{Quantum communication}. For each step $j \in [1,n]$, Alice and Bob perform the following steps.
    \begin{itemize}
        \item Alice and Bob draw a value from the random variables $X_j,Y_j\in\{0,1,2\}$ given by the probabilities $((1 - p_{\mathrm{v}})/2,(1- p_{\mathrm{v}})/2,p_{\mathrm{v}})$, and prepare the states $\rho^{A'_j}_{x} $, $ \rho^{B'_j}_{y}  $ corresponding to the ideal inputs $\left\{\ket{\gamma^{A'}},\ket{-\gamma^{A'}},\ket{\oslash} \right\}$, $\left\{\ket{\gamma^{B'}},\ket{-\gamma^{B'}},\ket{\oslash} \right\}$.
        \item Alice and Bob send their preparations through their corresponding quantum channels, which are manipulated and measured by a relay which implements a beam splitter and two threshold detectors.
    \end{itemize}
    \item \textit{Classical postprocessing}. For each round $j \in [n]$, Alice and Bob perform the following steps.
    \begin{itemize}
        \item The relay announces a value $C_j \in \{0,1,\bot\}$ corresponding to a zero click, a one click and either a double- or a no-click.
        \item Alice and Bob disclose their inputs for the round if either $C_j=\bot$, $X_j=2$, $Y_j=2$. Then, Alice further assigns a public register $I_j=1$. 
        \item Otherwise, Alice sets $I_j=0$ with probability $\pkey$ and $I_j=1$ with probability $1-\pkey$, revealing the inputs only if $I_j=1$.This probabilistic assignment is performed after the relay outcome $C_j$ is known, so that $\pkey$ represents the conditional probability of retaining an eligible round for key generation, rather than a fixed fraction of all protocol rounds.
    \end{itemize}
    \item \textit{Parameter estimation} Alice sets a raw  key register with $S_j = X_j$ when her input remains undisclosed, and $S_j = \bot$ else. Given the public information, Alice announces a length $\ell$ for the size of the final key, which also determines a length for the error correction register.
    \item \textit{Local randomization}. After parameter estimation, for each round $j$ retained for key generation, Alice independently generates a random bit
    $F_j\sim\mathrm{Bern}(p_{\mathrm{LR}})$, with $p_{\mathrm{LR}}\in[0,1/2]$, which is kept private and independent of Eve. She then applies the
    bit flip to her raw-key register according to $S_j \longleftarrow X_j\oplus F_j$.
    \footnote{Performing local randomization after parameter estimation avoids consuming private randomness on testing and discarded rounds, thereby reducing the number of random bits required from a QRNG. This may be advantageous in practical implementations where random-number generation constitutes an intensive resource.}
    \item \textit{Information reconciliation}. Alice and Bob perform a one-way, direct reconciliation error correction to remove the disagreements between their secret keys. Bob obtains a guess of Alice's keys and  exchanges the hashed key with her. If the hashes do not agree, the key length is set to zero.
    \item \textit{Privacy amplification}. Alice and Bob generate the shorter but fully private final key via a privacy amplification subroutine, such as another hash function.
\end{itemize}

Given these steps, we define the complete parameter estimation register $\hat{C}$ with values in the alphabet
\begin{align}
    \hat{\mc{C}} = \{(\bot,\bot)\times(0,1)\} \cup \{(0,1,2)\times (0,1,2)\times(0,1,\bot)\}
\end{align}
In particular, we note that the register $I$ is fully determined by $\hat{C}$, as it can be only ambiguous if the inputs are not revealed, but the cases $\hat{C} = (\bot,\bot)\times(0,1)$ and $\hat{C} = (0,1) \times (0,1)\times(0,1)$ already indicate $I=0$ and $I=1$ respectively. For future reference, let us also define
\begin{align}
    \tilde{\mc{C}} = \hat{\mc{C}}\backslash \{(\bot,\bot)\times(0,1)\}.
\end{align}
which is the alphabet of genuine parameter estimation symbols. 

For the rest of our discussion, let us switch from the initial, prepare-and-measure approach into its corresponding entanglement-based picture, which is fully equivalent thanks to the source-replacement scheme \cite{BBM92,Curty2004}. Writing $p_{xy} := p_A(x)p_B(y)$ for the joint probability of preparation, we have that the initial state at round $j \in [n]$ is, in the entanglement-based image,

\begin{align}\label{eq:InitialEBState}
    \rho_{[AA'BB']_j} = \sum_{x,y=0}^2 \sqrt{p_{xy}} \ket{x,\rho^{A'}_x}_{[AA']_j} \otimes \ket{y,\rho^{B'}_y}_{[BB']_j}
\end{align}

Comparing the state \eqref{eq:InitialEBState} with the preparations outlined in the protocol description, we observe that \eqref{eq:InitialEBState} is actually given by pure states. This notion is valid as we can consider that the corresponding purifications (also called shield systems \cite[Lemma 8.3]{tupkary2026rigorouscompletesecurityproof}) lie at the laboratories of Alice and Bob and are therefore never accessed by Eve. Hence, these purifications can in principle be incorporated as part of $[AB]_j$ to further constrain the state shared by the honest parties. We will omit this technicality here, as we eventually replace the inputs of Alice and Bob with a different set of reference states to model source imperfections (see Subsection \ref{Subsec:sourceImperfections}).

Quantitatively, we can describe the total map $\mathcal M_j:A_jB_jC_j\longrightarrow S_j\hat C_j$ of round no. $j$ by Alice, Bob and the repeater as
\begin  {align}
    \mc{M}_j(\cdot) &= \pkey \sum_{x,y,c=0,1} \Tr[\pro{x}_{A_j} \otimes \pro{y}_{B_j} \otimes \pro{c}_{C_j} (\cdot)] \nonumber \\
    &\qquad \times [
        (1-p_{\mathrm{LR}})\pro{x}_{S_j}
        + p_{\mathrm{LR}}\pro{1-x}_{S_j}
    ] \otimes \pro{\bot \bot,c}_{\hat{C}_j} \nonumber \\
    & \quad + \sum_{x,y=0,1,2} \Tr[\pro{x}_{A_j} \otimes \pro{y}_{B_j} \otimes \pro{\bot}_{C_j} (\cdot)] \pro{\bot}_{S_j} \otimes \pro{x,y,\bot}_{\hat{C}_j} \nonumber \\
    & \quad + \sum_{y=0,1,2}\sum_{c = 0,1} \Tr[\pro{2}_{A_j} \otimes \pro{y}_{B_j} \otimes \pro{c}_{C_j} (\cdot)] \pro{\bot}_{S_j} \otimes \pro{2,y,c}_{\hat{C}_j} \nonumber \\  
    & \quad + \sum_{x=0,1}\sum_{c = 0,1} \Tr[\pro{x}_{A_j} \otimes \pro{2}_{B_j} \otimes \pro{c}_{C_j} (\cdot)] \pro{\bot}_{S_j} \otimes \pro{x,2,c}_{\hat{C}_j} \nonumber \\  
    & \quad + (1-\pkey) \sum_{x,y,c=0,1} \Tr[\pro{x}_{A_j} \otimes \pro{y}_{B_j} \otimes \pro{c}_{C_j} (\cdot)] \pro{\bot}_{S_j} \otimes \pro{x,y,c}_{\hat{C}_j} \label{eq:ExpandedTotalMap} \\
    &:= \pkey \mkey(\cdot)_{{S_j} \hat{C}_j} + \pro{\bot}_{S_j} \otimes \mtest(\cdot)_{\hat{C}_j}  \label{eq:TotalMap}
\end{align}
Where in the last line we defined the subnormalized maps for the key and testing rounds. Observe that the maps \(\mathcal M_j\) are identical up to the labeling of their input and output registers, in such a way that from now onwards we will use the symbol $\mathcal M$ to refer to the single round map. 
Importantly, Eq.~\eqref{eq:TotalMap} implements an outcome-dependent probabilistic selection of key and testing rounds. A round is retained for key generation with probability $\pkey$ only when $X_j,Y_j,C_j\in\{0,1\}$. Consequently, the overall key-generation probability depends on the relay statistics and may therefore be influenced by Eve's attack. Unlike previous numerical MEAT implementations assuming a fixed overall testing probability \cite{kamin25MEATsecurity,tupkary2026rigorouscompletesecurityproof,wiesemann2026}, our single-round map explicitly incorporates this dependence. Furthermore, this approach reduces the consumption of private randomness, since the probabilistic key/testing assignment requires QRNG resources only for eligible rounds, rather than for every protocol round.
The total protocol is provided by the state after \(n\) concatenations of single-round actions, as described in the protocol outline, that can equivalently be expressed as the tensor product \(\mathcal M^{\otimes n}\)
\begin{equation}
\nu_{[S\hat C]_1^nE}
=
\left(
\mathcal M^{\otimes n}
\otimes\mathbb I_{E}
\right)
\circ
\left(
\mathbb I_{[AB]_1^n}
\otimes\Lambda
\right)
\left[
\rho_{AA'BB'}^{\otimes n}
\right].
\label{eq:totalState}
\end{equation}
where $\Lambda: [A'B']_1^n \to[C]_1^n E$ represents the general attack by Eve on all signals, incorporating the untrusted receiver's measurements. In particular, Eve's quantum side information (e.g. via quantum memories) and all possible correlations across rounds are fully accounted for by the global attack channel $\Lambda$, with $E$ representing Eve's quantum register.

For the security analysis, we mention the need for an auxiliary register $R$ that provides a coherent representation of the classical key register $S$. This register is used for the mathematical construction of the key and pinching maps (whose explicit definition is provided in Appendix \ref{App:NoisyPreproc}), and the local randomization.
Given the definition of the steps for all rounds and the construction of the total state, we are in conditions to develop the security proof and describe quantitatively the secret key rate.

\section{Finite-size analysis}\label{sec:FiniteSize}

Following the approach of \cite[Theorem 12]{kamin25MEATsecurity}, we can use the total state \eqref{eq:totalState} after the quantum communication to derive the expression for the finite secret key in terms of single-round quantities.

\begin{remark}
    We note that the MEAT is actually only well-defined when all the involved registers are finite-dimensional. This is not a problem in our case, as the set of inputs are finite-dimensional, and the outputs of the quantum relay (i.e. register $C$) are given by a finite-dimensional register. As a result, only Eve's register can be infinite-dimensional. For such case, one can safely apply the MEAT assuming a finite dimension for Eve, and extend the bound for the infinite-dimensional case using \cite[Appendix A]{tupkary2026rigorouscompletesecurityproof}.
\end{remark}

Unlike the more general framework of
Ref.~\cite[Sec.~4]{tupkary2026rigorouscompletesecurityproof},
we do not consider on-the-fly public announcements,
where classical information from earlier rounds may become
available to Eve before the quantum communication is completed.
Instead, all public announcements are performed after the
transmission and measurement of the $n$ quantum signals.
This restriction simplifies the security analysis, since the
classical postprocessing can be represented by a single-round
map $\mathcal M$, identical for every round \cite[Corollary 4.2]{arqand2025MEAT}, yielding the
overall map $\mathcal M^{\otimes n}$ as in
Eq.~\eqref{eq:totalState}. Consequently, no additional
treatment of announcement timings is required, while
security against general coherent attacks is preserved
through the unrestricted global attack channel $\Lambda$.
Under these assumptions, applying
\cite[Theorem 8.1]{tupkary2026rigorouscompletesecurityproof}
yields an $\epsilon_\mathrm{PA}$-secret and
$\epsilon_\mathrm{EC}$-correct implementation of the
protocol, with a secret-key length given by

\begin{align}\label{eq:SKR}
    \ell(\hat{c}_1^n) = \max \left\{ 0, \left\lfloor \hat{f}_{\text{full}}(\hat{c}_1^n) - \lambda_{\text{EC}}(\hat{c}_1^n) - \left\lceil \log\left(\frac{1}{\epsilon_{\mathrm{EC}}}\right) \right\rceil - \frac{\alpha}{\alpha - 1} \log\left(\frac{1}{\epsilon_{\mathrm{PA}}}\right) + 2 \right\rfloor \right\}
\end{align}
where we used the full tradeoff function
\begin{align}\label{eq:fulltradeoff}
    \hat f_{\mathrm{full}}(\hat c_1^n)
    :=
    n\left(
    \sum_{\hat c\in\hat{\mc C}}
    f(\hat c)F^{\mathrm{obs}}(\hat c)
    +\kappa
    \right),
\end{align}
according to the observed, finite frequency vector
\begin{align}
     F^{\mathrm{obs}}(\hat c)
    :=
    \frac{1}{n}
    \sum_{j=1}^{n}
    \mathbf{1}_{\{\hat c_j=\hat c\}},
    \qquad
    \hat c\in\hat{\mc C}.
    \label{eq:Fobs}
\end{align}

On the other hand, following ~\cite[Lemma~8.8]
{tupkary2026rigorouscompletesecurityproof},
we define $\kappa$ through a marginal-state
optimization that provides a lower bound on the single-round MEAT entropy
contribution. The resulting program reads
(hereafter we use the simplified notation
$\mathcal M\otimes\mathbb I_{\tilde E}
\equiv\mathcal M$)

\begin{equation}
\begin{gathered}
  \kappa :=
  \inf_{\sigma \in \Sigma}
  H_{\alpha}^{\uparrow,f}
  (S | \hat{C} \tilde{E} )_\sigma \\
\text{s.t. } \\
\Sigma :=
\left\{
{\mc{M}[}\mathrm{Pur}
\left( \omega_{ABC } \right)]
\mid
\omega \in \mc{D}(ABC),
\, \omega_{AB} = \rho_{AB}
\right\},
\end{gathered}
\end{equation}
with $\alpha \in (1,2)$,
$f(\hat{c}): \hat{\mc{C}} \to \mathbb{R}$
a tradeoff function, which can be arbitrarily
chosen but can be optimized as shown in
Subsection~\ref{subsec:tradeoff},
and 
$\mathrm{Pur}$ a purifying function of $ABC$ on $\tilde E$, for $\tilde E$ an auxiliary register of sufficient
dimension to purify any compatible single-round
state $\omega_{ABC}$.
In contrast, we recall that $E$ denotes Eve's side information
resulting from the global attack $\Lambda$. As established in
Ref.~\cite[Lemma~8.8]
{tupkary2026rigorouscompletesecurityproof},
this marginal formulation provides a valid
lower bound on the original optimization
over Eve's attack.

\begin{remark}
    Notice that the purification $\tilde{E}$ concerns the state before the application of the local postprocessing
map $\mathcal M$, which includes Alice's trusted
local randomization. Thus, while $\tilde E$
accounts for all possible adversarial correlations
prior to postprocessing, the private randomness
subsequently introduced by Alice's bit-flip
operation remains inaccessible to Eve.
\end{remark}

We now proceed to characterize the set of compatible single-round states. Since the register $C$ is classical, the marginal state $\omega_{ABC}$ admits a block-diagonal decomposition with respect to the possible relay announcements.
Moreover, by the data processing inequality
\cite[Lemmas~6.2 and~6.8]
{tupkary2026rigorouscompletesecurityproof},
the minimization over purifications can be
expressed as a minimization over all extensions
$\omega_{ABC\tilde E}$ of the compatible
marginal states. Thus we find
\begin{equation}
\begin{gathered}
\kappa := \inf_{\omega_{ABC\tilde E}}
H_{\alpha}^{\uparrow,f}
(S|\hat C\tilde E)_
{\mc{M}(\omega)}
\\
\text{s.t.}\\
\omega_{ABC\tilde E}\in\mc{D}(ABC\tilde E),\\
\omega_{ABC}
= \bigoplus_{c\in\{0,1,\bot\}}\tilde{\omega}^c_{AB},
\quad \tilde{\omega}^c_{AB}\succeq 0,\\
\sum_{c\in\{0,1,\bot\}}\tilde{\omega}^c_{AB}
=\rho_{AB}.
\end{gathered}
\label{eq:kappa_def}
\end{equation}
Here, we particularly remark how we have applied a decomposition of the state $\omega_{ABC}$ as a consequence of being $C$ classical, albeit leading to a combination of states $\{\tilde{\omega}^c_{AB}\}_{c=0,1,\bot}$ whose probabilities are unknown.

Let us now proceed with the simplification for the objective function. As a first step, we use the definition of $f-$weighted Rényi entropy \cite{arqand2025MEAT} together with the description of the postprocessing map \eqref{eq:TotalMap}
\begin{align}
    H_{\alpha}^{\uparrow,f} (S | \hat{C}  \tilde{E} )_{\mc{M}(\omega)} &= \frac{\alpha}{1-\alpha} \log \left(
    \sum_{{\hat c} \in \tilde{\mathcal{C}}} \Tr[\mtest(\omega)_{\hat c}] 2^{\frac{\alpha-1}{\alpha}f ({\hat c})} \right. \nonumber \\
    & \left.+ \pkey \sum_{c'=0,1} \Tr[\mkey(\omega)_{|\bot,\bot,c'}] 2^{\frac{1-\alpha}{\alpha} (-f ({\bot,\bot,c'})
    + H^\uparrow_\alpha (S|\tilde{E})_{\mkey(\omega)_{|\bot,\bot,c'}})} \right).
\end{align}

In particular, we have that $\hat c \in \tilde{\mc{C}}$ induces $S = \bot$ deterministically, such that the corresponding entropy is zero. For rounds spent on the key distillation, we further simplify the conditional sandwiched Rényi entropy as

\begin{align} H^\uparrow_\alpha (S|\tilde{E})_{\mkey(\omega)_{|\bot,\bot,c'}} &\geq
-{H}^\downarrow_\mu (S| RAB)_{\mkey(\omega)_{|\bot,\bot,c'}}\label{eq:21} \\ 
&\geq {D}_\mu\left(\mathcal{G}^{c'}(\tilde{\omega}^{c'})\,\middle\|\,\mathcal{Z\circ G}^{c'}(\tilde{\omega}^{c'}) \right) \label{eq:23} \\
&= \frac{1}{\mu-1} \log\left( \frac{\hat{\Psi}_\mu^{c'}(\tilde{\omega}^{c'})}{\Tr[\mkey(\omega)_{|\bot,\bot,c'} ]} \right), \label{eq:24}
\end{align}
with $\mu = 1/\alpha$, and $\hat{\Psi}^{c'}_\mu (\cdot) := {\Psi}_\mu (\g^{c'}(\cdot),\zg^{c'}(\cdot))$. This entropy bound is derived via the  introduction of an auxiliary register $R$, such that the key register $S$ is then obtained by measuring $R$ in the computational basis (see Appendix \ref{App:NoisyPreproc}). Inequality \eqref{eq:21} holds from the entropic duality relations for sandwiched Rényi entropies  \cite{Tomamichel2014}. For equations \eqref{eq:23} and \eqref{eq:24}, we have followed the steps of \cite[Appendix A]{Navarro2025} which introduce the key and pinching maps $\g$, $\z$ involved on a key distillation round together with \eqref{eq:QRenyiDiv}. 
With this lower bound we find

\begin{align}
    H_{\alpha}^{\uparrow,f} (S |\hat C \tilde{E} )_{\mc{M}(\omega)} &\geq \frac{\alpha}{1-\alpha} \log \left( \sum_{{\hat c} \in \tilde{\mathcal{C}}} \Tr[\mtest(\omega)_{\hat c}] 2^{\frac{\alpha-1}{\alpha}f ({\hat c})} \right. \nonumber \\
    & \qquad  \left.+ \pkey \sum_{c'=0,1} 2^{\frac{\alpha-1}{\alpha} f ({\bot,\bot,c'})} \hat{\Psi}^{c'}_\mu(\tilde{\omega}^{c'}) \right).
\end{align}
Using this simplification, and thanks to the monotonicity of the negative logarithm, we can perform the optimization in \eqref{eq:kappa_def} by only maximizing the argument of said logarithm. To conclude, we use the FastRényiQKD cone \cite{Navarro2025}
\begin{equation}\label{eq:fast_qkd_cone}
    \mc{K}^{\mu, \gmap, \zgmap, \widehat S}_\text{FastRényiQKD} = \left\{(u,\rho) \in \mathbb{R} \times \mathbb H^{d}_\succ; u \ge - \bar{{\Psi}}_\mu(\rho) \right\}\,,
\end{equation}
to numerically model the optimization with respect to the Rényi divergences, with $\gmap,\zgmap$ representing the $\g,\zg$ maps after performing a facial reduction (leading to an isometry $\widehat S$ that encodes said reduction), which will be explicitly provided in Section \ref{Sec:numerics}. As a result of this reduction, the original function $\hat{\Psi}_\mu (\cdot)$ transforms into $\bar{\Psi}_\mu(\cdot)$.

\begin{remark}
    We note that the FastRényiQKD cone appears as a result of relaxing the original optimization problem, and that one could perform the full optimization corresponding to the true Rényi cone \cite{Navarro2025}. However, this comes at a costly time overhead, while leading to a residual improvement according to a factor $\mu$ \cite{schmitt2026marginalprettygood}, which in our case lies in the order $\mu -1 \lesssim 10^{-4}$ for all relevant block sizes.
\end{remark}

As a result, we arrive at the optimization
\begin{equation}\label{eq:kappaMax}
\begin{gathered}
  \max_{\omega, \{u^{c'}\}_{c'}}  \sum_{{\hat c} \in \tilde{\mathcal{C}}}\Tr[\mtest(\omega)_{\hat c}] 2^{\frac{\alpha-1}{\alpha}f ({\hat c})}  -  \pkey \sum_{c'=0,1} 2^{\frac{\alpha-1}{\alpha} f ({\bot,\bot,c'})} u^{c'} \\
\text{s.t. } \\
\omega_{ABC} = \bigoplus_{c=0}^2 \tilde{\omega}^c_{AB}, \; \tilde{\omega}^c_{AB} \succeq 0\\
\sum_{c=0}^2 \tilde{\omega}^c_{AB} =\rho_{AB} \\
(u^{c'},\tilde{\omega}^{c'}_{AB}) \in \mc{K}^{\mu, \gmap^{c'}, \zgmap^{c'}, \widehat S^{c'}}_\text{FastRényiQKD}, \, c' = 0,1.
\end{gathered}
\end{equation}

\subsection{Source imperfections}\label{Subsec:sourceImperfections}

The optimization in \eqref{eq:kappaMax} presupposes that Alice and Bob know exactly the preparation for their states as they fix the marginal of the state on $AB$. Albeit this condition is valid since registers $[AB]_1^n$ are never accessible to Eve, it also assumes wrongfully that the characterization of the states sent through the quantum channel is absolute.

This  demand is unrealistic as imperfections, such as in the phase and amplitude modulation, can induce deviations and eventually open side channel attacks through uncontrolled degrees of freedom \cite{BSI_Report}. Hence, it is desirable to incorporate such imperfect characterization for the sources in the security analysis, and bound any information leakage.

We include this feature in our model by following the partial source characterization of Ref. \cite{pereira2025optimalkeyratesquantum}. I.e., we substitute the exact knowledge of the marginal by a fidelity bound with respect to a set of known reference states, and we apply it in the calculation of \eqref{eq:kappaMax}.

\subsubsection{Partial state characterization}

We recall that $\rho^{A'}_x$ denotes the state sent by Alice at round no. $j$ given by a value $X = x \in \{0,1,2\}$, and idem for Bob with $\rho^{B'}_y$.  For sake of simplicity, since we are assuming identical preparations at each round, we drop the dependence on the round number $j.$ Let now $\left\{\ket{\phi^{A'}_x}\right\}_{x=0}^{2}$ (respec. for Bob) be known pure reference states, i.e. the nominal preparations $\left\{\ket{\gamma^{A'}},\ket{-\gamma^{A'}},\ket{\oslash}\right\}$ (respec. for Bob) of the protocol description, including any trusted imperfections such as a deterministic phase offset. 

Next, we combine the uncorrelated sources by Alice and Bob by considering their joint emission $\rho_{xy} = \rho^{A'}_x \otimes \rho^{B'}_y$, and define the joint reference states
\begin{align}\label{eq:jointReference}
    \ket{\phi_{xy}} := \ket{\phi^{A'}_x} \otimes \ket{\phi^{B'}_y}, 
\end{align}
to constrain the source via the joint fidelity bounds given tolerance coefficients $\varepsilon^{A}_{x}, \varepsilon^{B}_{y} \in (0,1)$
\begin{align}\label{eq:jointFidelity}
    \bra{\phi_{xy}}\rho_{xy}\ket{\phi_{xy}} \geq (1 - \varepsilon^{A}_{x})(1 - \varepsilon^{B}_{y}), \qquad \forall x,y \in \{0,1,2\}.
\end{align}

We note that these inequalities place no restrictions on the Hilbert space where the deviations live. Any auxiliary degree of freedom (together with its corresponding side channel attack) is included in the security analysis as long as it does not push the emitted state further than $\varepsilon-$away from the reference in fidelity (which results in a protocol abortion). 

Furthermore, the fidelity bounds set for Alice and Bob can be observed experimentally by a finite set of measurements \cite{Kliesch2021}, such that \eqref{eq:jointFidelity} becomes a reliable certificate to bound source imperfections. Within our context, and for the sake of simplicity, let us merge these coefficients according to their joint value as
\begin{align}\label{eq:simplifyepsilons}
    (1 - \varepsilon^{A}_{x})(1 - \varepsilon^{B}_{y}) \to (1 - \varepsilon_{x,y}).
\end{align}

Now, we can use \cite[Lemma 1]{pereira2025optimalkeyratesquantum} to reduce the set of states satisfying \eqref{eq:jointFidelity} to a family of pure states.

\begin{proposition} \cite{pereira2025optimalkeyratesquantum,Curras2025numerical}\label{lem:pureReduction}
    Let $\{\rho_{xy}\}_{x,y=0}^2$ satisfy \eqref{eq:jointFidelity}. Then there exists a CPTP map $\Phi$ (also called source map) and pure states $\left\{\ket{\psi_{xy}}\right\}_{x,y=0}^2$ verifying
    \begin{align}\label{eq:pureForm}
        \ket{\psi_{xy}} = \sqrt{1-\varepsilon_{xy}}\ket{\phi_{xy}} + \sqrt{\varepsilon_{xy}}\ket{\phi^{\perp}_{xy}}, \qquad \braket{\phi^{\perp}_{xy}}{\phi_{xy}} = 0,
    \end{align}
    such that $\rho_{xy} = \Phi \left(\pro{\psi_{xy}} \right)$ for all $x,y=0,1,2$.
\end{proposition}

From an operational perspective, preparing $\rho_{xy}$ is equivalent to preparing $\ket{\psi_{xy}}$ and then applying the source map. As $\Phi$ acts only on systems that are sent through the quantum channels, we can give it to Eve in order to lower bound the protocol performance. Formally, we can consider that Alice and Bob prepare an ideal, entanglement-based state $\psi^{\otimes n}_{AA'BB'}$, then apply $n$ copies of $\Phi$ on said ideal inputs to generate the imperfect, real copies -- therefore, the total state after the quantum communication \eqref{eq:totalState} becomes 

\begin{equation}
\nu_{[S\hat C]_1^nE}
=
\left(
\mathcal M^{\otimes n}
\otimes\mathbb I_{E}
\right)
\circ
\left(
\mathbb I_{[AB]_1^n}
\otimes\Lambda
\right)
\left[
(\mathbb I_{[AB]_1^n}\otimes\Phi^{\otimes n})\left(
\psi_{AA'BB'}^{\otimes n}
\right)
\right].
\end{equation}

The security of the protocol resulting from this implementation is equivalent to the one using ideal states by absorbing $\Phi^{\otimes n}$ into $\Lambda$ thanks to\footnote{In fact, \cite[Lemma 9.1]{tupkary2026rigorouscompletesecurityproof} is much more general as it does not require, for instance, a tensor product between the source maps.} \cite[Lemma 9.1]{tupkary2026rigorouscompletesecurityproof}.

\subsubsection{Marginal states under the partial characterization}

The logic of this new version of the protocol, where Alice and Bob use virtual preparations, is connected to the real one via the marginal constraint. Provided our imperfect sources, the entangled state coming from the states \eqref{eq:pureForm} induces a new marginal for Alice and Bob
\begin{align}
    \psi_{AB} &= \Tr_{A'B'}\!\left[\pro{\psi}_{AA'BB'}\right] \\
    &= \sum_{x,y,x',y'} \sqrt{p_{xy}p_{x'y'}}\braket{\psi_{x'y'}}{\psi_{xy}}\ket{xy}\bra{x'y'}_{AB}. 
    \label{eq:marginalImperfect}
\end{align}
Expanding the overlaps via \eqref{eq:pureForm} and collecting the two branches with the weights

\begin{align}\label{eq:weights}
    w_{xy 0} := \sqrt{1-\varepsilon_{xy}} && w_{xy 1} := \sqrt{\varepsilon_{xy}}
\end{align}
and using the shorthand notation
\begin{align}
    \ket{\phi^{(0)}_{xy}} := \ket{\phi_{xy}} && \ket{\phi^{(1)}_{xy}} := \ket{\phi^{\perp}_{xy}}
\end{align}
the states can be written
\begin{align}
    \ket{\psi_{xy}} = \sum_{a=0,1} w_{xya}\ket{\phi^{(a)}_{xy}}.
\end{align}
The overlaps induce the inner product
\begin{align}\label{eq:overlapExpansion}
    \braket{\psi_{x'y'}}{\psi_{xy}} = \sum_{a,b = 0,1} w_{x'y'a}\, w_{xy b}\; G_{(x'y',a),(xy,b)}.
\end{align}
Here we have introduced the Gram matrix \cite{pereira2025optimalkeyratesquantum} of the total set of reference states, given by the matrix elements

\begin{align}\label{eq:GramDef}
    G_{(xy,a),(x'y',b)} := \braket{\phi^{(a)}_{xy}}{\phi^{(b)}_{x'y'}}, \qquad x,y,x',y' \in \{0,1,2\} \wedge   a,b \in \{0,1\}.
\end{align}
By construction, $G \in \mathbb{H}^{18}_ \succeq$. The Gram matrix has known entries in the reference states as
\begin{subequations} \label{eq:GramConstraints}
\begin{align}
    G_{(xy,0),(x'y',0)} &= \braket{\phi_{xy}}{\phi_{x'y'}}, \qquad  x,y,x',y' \in \{0,1,2\} \label{eq:GramKnownBlock}\\
    G_{(xy,1),(xy,1)} &= 1, \;\;\, \qquad \qquad\qquad x,y \in \{0,1,2\} \label{eq:GramNorm}\\
    G_{(xy,0),(xy,1)} &= 0, \;\;\, \qquad\qquad\qquad x,y \in \{0,1,2\} \label{eq:GramOrth}
\end{align}
\end{subequations}
namely the fully characterized upper-left block \eqref{eq:GramKnownBlock}, the normalization of the unknown deviations \eqref{eq:GramNorm}, and their orthogonality with respect to the corresponding reference \eqref{eq:GramOrth}. All other entries (under the condition of positive semidefiniteness) are chosen by Eve, and must be optimized in the final program.

\begin{remark}\label{rem:productStructure}
    This characterization for imperfect sources is actually a relaxation: the unknown vector $\ket{\phi^{\perp}_{xy}}$ in \eqref{eq:pureForm} is not necessarily separable  in $A'B'$. Therefore, we are allowing Eve to correlate between the two states despite the fact that their preparations are known to be independent.

    Ideally, one could impose individual bounds on the preparations of Alice and Bob, instead of a joint bound  \eqref{eq:jointFidelity}. However, enforcing a separability $\rho_{xy}=\rho^{A}_x\otimes\rho^{B}_y$ would require $G$ to be built as a product of two smaller Gram matrices, which does not yield a convex constraint for the final program.
\end{remark}

In order to simplify our formulation for the marginal, we define the admissible Gram set according to the no. of inputs $n_A,n_B$ by Alice and Bob
\begin{align}\label{eq:GramSet}
    \mathfrak{G} := \left\{ G \in \mathbb{H}^{2n_An_B} \;\middle|\; G \succeq 0 \ \text{ and } \ \eqref{eq:GramConstraints} \text{ hold} \right\},
\end{align}
which is the intersection of the positive semidefinite cone with an affine subspace, therefore the set is closed and convex. Second, we denote $\Xi(\cdot)$  as the linear map that reconstructs the marginal state from the Gram matrix
\begin{align}\label{eq:XiMap}
    \Xi[G] := \sum_{x,y,x',y'} \sqrt{p_{xy}p_{x'y'}} \left( \sum_{a,b=0,1} w_{x'y'a}\, w_{xy b}\, G_{(x'y',a),(xy,b)} \right) \ket{xy}\bra{x'y'}_{AB}.
\end{align}
Combining \eqref{eq:marginalImperfect} with \eqref{eq:overlapExpansion}, the statement of this section is compactified to
\begin{align}\label{eq:marginalAsMap}
    \psi_{AB} = \Xi\left[G\right], \qquad G \in \mathfrak{G}.
\end{align}
The particular case of a perfectly characterized source is recovered by setting $\varepsilon_{xy}=0$, such that all terms involving $a=1$ or $b=1$ at \eqref{eq:XiMap} vanish and $\Xi[G]$ reduces to a single state.

Since each admissible Gram matrix $G$ defines
a pure virtual source-replacement state
with marginal $\Xi[G]$, the marginal-state
reduction of ~\cite[Lemma~8.8]
{tupkary2026rigorouscompletesecurityproof}
remains applicable for every fixed
$G\in\mathfrak G$.
To account for the partial source
characterization, we further optimize over
all admissible Gram matrices, replacing
the fixed marginal condition
$\omega_{AB}=\rho_{AB}$ in
Eqs.~\eqref{eq:kappa_def} and
\eqref{eq:kappaMax} with
$\omega_{AB}=\Xi[G]$.
We thus arrive at the following program
\begin{equation}
\begin{gathered}
  \max_{\omega,\, G,\, \{u^{c'}\}_{c'}}  \sum_{{\hat c} \in \tilde{\mathcal{C}}} \Tr[\mtest(\omega)_{\hat c}] 2^{\frac{\alpha-1}{\alpha}f ({\hat c})}  -  \pkey \sum_{c'=0,1} 2^{\frac{\alpha-1}{\alpha} f({\bot,\bot,c'})} u^{c'} \\
\text{s.t. } \\
\omega_{ABC} = \bigoplus_{c=0}^2 \tilde{\omega}^c_{AB}, \; \tilde{\omega}^c_{AB} \succeq 0\\
\sum_{c=0}^2 \tilde{\omega}^c_{AB} = \Xi\left[G\right], \qquad G \in \mathfrak{G} \\
(u^{c'},\tilde{\omega}^{c'}_{AB}) \in \mc{K}^{\mu, \gmap^{c'}, \zgmap^{c'},  \widehat S^{c'}}_\text{FastRényiQKD}, \, c' = 0,1.
\end{gathered}
\label{eq:kappaMaxImperfect}
\end{equation}
Which is again a conic program, as the optimization with respect to the Gram matrix only involves a semidefinite constraint and equalities. In particular, we observe that the trace-one condition, previously encoded in the ideal marginal, can also be retrieved from the Gram matrix via (\ref{eq:GramKnownBlock}-\ref{eq:GramOrth}).

\subsection{Tradeoff functions}\label{subsec:tradeoff}

As a last step, we have that \eqref{eq:kappaMax} requires to set first a tradeoff function $f$. This can be optimally chosen based on the fixed-length version of the security proof (\cite[Lemma 4.12]{arqand2025MEAT} and \cite[eq (solve for dual)]{kamin25MEATsecurity}). Namely, we define a reference probability distribution $q_{\hat{C}}$ (see Sec.~\ref{Sec:numerics}), which models the performance of the protocol in an honest implementation, and formulate

\begin{equation}\label{eq:tradeoffProgram}
\begin{gathered}
  \min_{ \omega, G,\hat\lambda} \frac{\alpha}{\alpha-1} D_{\mathrm{KL}}(\hat\lambda|| \mc{M}(\omega)_{\hat{C}} )  + \sum_{\hat c \in \hat{\mc{C}}} \hat\lambda(\hat c) H_{\alpha}^{\uparrow} (S |\tilde  E )_{\mc{M}(\omega)_{|\hat c}}, \\
\text{s.t. } \\
\omega_{ABC} = \bigoplus_{c=0}^2 \tilde{\omega}^c_{AB}, \; \tilde{\omega}^c_{AB} \succeq 0\\
\sum_{c=0}^2 \tilde{\omega}^c_{AB} = \Xi\left[G\right], \qquad G \in \mathfrak{G} \\
\hat\lambda - q_{\hat C}=0.
\end{gathered}
\end{equation}
Then, the dual variables associated with the last constraint of the program will represent an optimal choice for $f$.

The formulation of \eqref{eq:tradeoffProgram} as a conic program follows a similar outline as for \eqref{eq:kappa_def}. First of all, we eliminate from the sum of entropies those terms where $S$ is fully determined by the classical conditioning
\begin{align}
    \sum_{c \in \hat{\mc{C}}} \hat\lambda(c) H_{\alpha}^{\uparrow} (S | \tilde{E} )_{\mc{M}(\omega)_{|c}} &= \sum_{c'=0,1} \hat\lambda(\bot,\bot,c') H_{\alpha}^{\uparrow} (S | \tilde{E})_{\mkey(\omega)_{|\bot,\bot,c'}}.
\end{align}
Next, we reduce the remaining entropies via dualization as in \eqref{eq:24}

\begin{align}
    H_{\alpha}^{\uparrow} (S |  \tilde{E} )_{\mkey(\omega)_{|\bot,\bot,c'}} &\geq \frac{1}{\mu -1} \log \left( \frac{ \hat{\Psi}^{c'}_\mu(\tilde{\omega}^{c'})}{\Tr[\mkey(\omega)_{|\bot,\bot,c'}]} \right) \\
    & = \frac{\alpha}{1-\alpha} \log \left( \frac{ \hat{\Psi}^{c'}_\mu(\tilde{\omega}^{c'})}{\Tr[\mkey(\omega)_{|\bot,\bot,c'}]} \right).
    \label{eq:51}
\end{align}
Then, we split the Kullback-Leibler divergence into the parameter estimation and key elements
\begin{align}
\label{eq:52}
    D_{\mathrm{KL}}(\hat\lambda|| \mc{M}(\omega)_{\hat{C}} )  = D_{\mathrm{KL}}(\tilde{\lambda}|| \mtest(\omega)_{\hat{C}} )  + \sum_{c'=0,1} \hat\lambda(\bot,\bot,c') \log \left( \frac{\hat\lambda(\bot,\bot,c') }{\pkey\Tr[\mkey(\omega)_{|\bot,\bot,c'}]} \right).
\end{align}
Where $\tilde{\lambda}$ is the component defined on $\tilde{\mc{C}}$. Using \eqref{eq:51} and \eqref{eq:52} inside \eqref{eq:tradeoffProgram}, we reach that 
\begin{align}
    \frac{\alpha}{\alpha-1} & D_{\mathrm{KL}}(\hat\lambda|| \mc{M}(\omega)_{\hat{C}} )  + \sum_{\hat c \in \hat{\mc{C}}} \hat\lambda(\hat c) H_{\alpha}^{\uparrow} (S | \tilde{E} )_{\mc{M}(\omega)_{|\hat c}}\\
    \ge & \frac{\alpha}{\alpha-1}\Bigg( D_{\mathrm{KL}}(\tilde{\lambda}|| \mtest(\omega)_{\hat{C}} ) \nonumber \\
    & + \sum_{c'=0,1} \hat\lambda(\bot,\bot,c') \left[ \log \left( \frac{\hat\lambda(\bot,\bot,c') }{\pkey\Tr[\mkey(\omega)_{|\bot,\bot,c'}]} \right) - \log \left( \frac{ \hat{\Psi}^{c'}_\mu(\tilde{\omega}^{c'})}{\Tr[\mkey(\omega)_{|\bot,\bot,c'}]} \right) \right] \Bigg)
    \label{eq:SplitBound}
\end{align}

In order to eventually find an expression that is amenable for conic programming, we element-wise merge the key components
\begin{align}
     \log \left( \frac{\hat\lambda(\bot,\bot,c') }{\pkey  \Tr[\mkey(\omega)_{|\bot,\bot,c'}]} \right) - \log \left( \frac{ \hat{\Psi}^{c'}_\mu(\tilde{\omega}^{c'})}{\Tr[\mkey(\omega)_{|\bot,\bot,c'}]} \right)= - \log \left( \frac{\pkey \hat{\Psi}^{c'}_\mu(\tilde{\omega}^{c'})}{\hat\lambda(\bot,\bot,c')} \right).
\end{align}
So the final objective function is 
\begin{equation}
    \frac{\alpha}{\alpha-1}\left( D_{\mathrm{KL}}(\tilde{\lambda}|| \mtest(\omega)_{\hat{C}} )  - \sum_{c'=0,1} \hat\lambda(\bot,\bot,c') \log \left( \frac{\pkey \hat{\Psi}^{c'}_\mu(\tilde{\omega}^{c'})}{\hat\lambda(\bot,\bot,c')} \right) \right)
\end{equation}

The resulting expression allows now a formulation with the relative entropy and logarithmic cones 
\begin{align}
    \mc{K}_\mathrm{RE} &= \cl\{(u,q,p) \in \mathbb{R} \times \mathbb{R}_>^d \times \mathbb{R}_>^d : u \geq D_\mathrm{KL}(q\|p)\},  \\ 
    \mc{K}_{\log} &= \cl\left\{ (u, v, w) \in \mathbb{R} \times \mathbb{R}_> \times \mathbb{R}_> : u \leq v \log\left(w/v\right) \right\},
\end{align}
together with the FastRényiQKD cone \eqref{eq:fast_qkd_cone} (see \cite{Navarro2025} for a full derivation of the method). As a result, we arrive at the conic formulation

\begin{equation}\label{eq:tradeoffConic}
\begin{gathered}
  \min_{\omega, h_\mathrm{RE}, \{h_\mathrm{QKD}^{c'}, u^{c'}\}_{c'}, G, \hat\lambda}  \frac{\alpha}{\alpha-1}\left[ h_\mathrm{RE} - \sum_{c'=0,1} h^{c'}_\mathrm{QKD} \right] \\
\text{s.t. } \\
\omega_{ABC} = \bigoplus_{c=0}^2 \tilde{\omega}^c_{AB}, \; \tilde{\omega}^c_{AB} \succeq 0\\
\sum_{c=0}^2 \tilde{\omega}^c_{AB} = \Xi\left[G\right], \qquad G \in \mathfrak{G} \\(h_\mathrm{RE},\tilde{\lambda}, \mtest(\omega)_{\hat{C}}) \in \mc{K}_{\mathrm{RE}} \\
(h^{c'}_\mathrm{QKD},\hat\lambda(\bot,\bot,c'),-u^{c'}) \in \mc{K}_{\mathrm{log}}, \, c' = 0,1. \\
(u^{c'},\tilde{\omega}^{c'}_{[AB]}) \in \mc{K}^{\mu, \gmap^{c'}, \zgmap^{c'}, \widehat S^{c'}}_\text{FastRényiQKD}, \, c' = 0,1. \\
\hat\lambda - q_{\hat C} = 0
\end{gathered}
\end{equation}

\subsection{Expected secret-key length}

Following \cite[Sec.~VI.A]{kamin25MEATsecurity}, we now evaluate the expected value of the secret-key length \eqref{eq:SKR} under the honest implementation of the protocol. The expectation of the observed frequencies is given by 
\begin{equation}
    \mathbb E [F^{\mathrm{obs}}(\hat c)] = q_{\hat C}(\hat c),
\end{equation}
where $q_{\hat C}$ denotes the honest reference
distribution,
introduced in Subsection~\ref{subsec:tradeoff}
and explicitly constructed in
Sec.~\ref{Sec:numerics}. Using the definition of the complete tradeoff function \eqref{eq:fulltradeoff}, by linearity of  the expectation 
\begin{align}
    \mathbb E[\hat f_{\mathrm{full}}(\hat c_1^n)]
     =
    n\left(
    \sum_{\hat c\in\hat{\mc C}}
    f(\hat c)q_{\hat C}(\hat c)
    +\kappa
    \right).
\label{eq:expected_tradeoff}
\end{align}
As in~\cite[Sec.~VI.A]{kamin25MEATsecurity},
we can further assume that the expected
error-correction leakage satisfies
\begin{align}
    \mathbb E[
    \lambda_{\mathrm{EC}}(\hat c_1^n)]
    \leq
    \lambda_{\mathrm{EC}}(q_{\hat C}),
    \label{eq:expected_EC}
\end{align}
where $\lambda_{\mathrm{EC}}(q_{\hat C})$ denotes an upper bound on the expected reconciliation leakage associated with the honest reference distribution.  Its explicit
evaluation is provided in
Subsection~\ref{subsec:EC_cost}. Combining Eqs.~\eqref{eq:SKR},
\eqref{eq:expected_tradeoff}, and
\eqref{eq:expected_EC}, we obtain 
 \begin{align}
    \mathbb E[\ell(\hat c_1^n)]
    \geq\;
    n\left(
    \sum_{\hat c\in\hat{\mc C}}
    f(\hat c)q_{\hat C}(\hat c)
    +\kappa
    \right)
    -\lambda_{\mathrm{EC}}(q_{\hat C})-
    \left\lceil
    \log\left(
    \frac{1}{\epsilon_{\mathrm{EC}}}
    \right)
    \right\rceil-
    \frac{\alpha}{\alpha-1}
    \log\left(
    \frac{1}{\epsilon_{\mathrm{PA}}}
    \right)
    +2.
    \label{eq:expected_SKR}
\end{align}
This expression provides a lower bound on the expected secret-key length used for the evaluation of the secret-key rate, namely the key ratio with respect to the number of rounds $n$.

\subsection{Error correction cost}
\label{subsec:EC_cost}

Finally, we quantify the length of the public error-correction communication, \(\lambda_{\mathrm{EC}}(q_{\hat{C}})\). Following the standard approach in QKD analysis, we assume that it is calculated according to the Slepian-Wolf \cite{SlepianWolf} bound:
\begin{equation}
\lambda_{\mathrm{EC}}(q_{\hat{C}})
\leq
f_\mathrm{EC} nH(S|YIC),
\end{equation}
where \(f_\mathrm{EC}\geq1\) is the reconciliation inefficiency, and the conditional entropy accounts for Alice's local randomization. Here both the expectation value and the Shannon entropy are referred to the honest implementation of the protocol derived in Appendix \ref{app:q_honest}, since error correction is always performed via the authenticated classical channel. For all numerical simulations, we set \(f_\mathrm{EC}=1.16\), unless explicitly stated otherwise.

\section{The horizontal quantum channel}\label{sec:Horizontal}

We now introduce the free-space channel model employed in our MDI QKD
implementation. In the MDI configuration, Alice and Bob transmit their
optical states towards the untrusted relay through two distinct free-space
links. The following derivation is given for a generic horizontal link and
can be independently applied to the Alice-relay and Bob-relay channels,
whose geometrical and optical parameters are not required, in general, to
be identical.

The model is intended to describe horizontal free-space links propagating
at an approximately constant altitude $h$ over a broad range of operating
conditions. At sufficiently low altitudes, atmospheric extinction and
turbulence may both affect the propagation, whereas their contributions
progressively decrease as the altitude increases. The atmosphere-free regime is recovered when atmospheric extinction and turbulence become negligible. Whenever atmospheric propagation is relevant, we restrict the description below to the weak-turbulence regime.

A key difference with respect to a standard fiber channel is that the
free-space channel generally does not have a fixed transmittance. The
fraction of optical power collected by the receiver fluctuates as a
consequence of random transverse displacements of the beam, induced by
atmospheric beam wandering and pointing errors
\cite{pirandola2021limits,Karakosta2025,Mannalath2026GEO}.
For a horizontal path of length $L$ at approximately constant altitude
$h$, we therefore write the instantaneous channel transmittance as
\cite[Eq.~(17)]{pirandola2021limits}
\begin{align}
    \eta_h(r)
    =
    \eta_{\ell,h}(h,L)\,
    \eta_{\mathrm{col}}(r),
    \label{eq:horizontal_transmittance}
\end{align}
where $\eta_{\ell,h}$ accounts for deterministic atmospheric extinction,
whereas $\eta_{\mathrm{col}}(r)$ is the collection efficiency and depends
on the instantaneous transverse displacement $r$ of the beam centroid
with respect to the center of the receiver aperture. Within our model, the stochasticity of
the horizontal channel is therefore encoded in the random displacement
$r$, which induces a fluctuating instantaneous transmittance.

We stress that the absence of atmospheric turbulence does not, by itself,
make the horizontal channel deterministic. Even for propagation
essentially outside the atmosphere, finite pointing accuracy may randomly
displace the beam centroid with respect to the receiver aperture and thus
produce a fluctuating collection efficiency. The channel becomes
deterministic only in the additional ideal limit in which jittering
fluctuations are also neglected.

When atmospheric propagation is considered, a fundamental quantity
characterizing turbulence is the refractive-index structure constant
$C_n^2$. Its altitude dependence can be described, for instance, by the
Hufnagel--Valley model
\cite{Hufnagel1974,Valley1980,AndrewsLaserBeam},
\begin{align}
    C_n^2(h)
    &=
    5.94\times10^{-53}
    \left(
        \frac{v}{27}
    \right)^2
    h^{10}e^{-h/1000}
    +2.7\times10^{-16}e^{-h/1500}
    +Ae^{-h/100},
    \label{eq_ref_index_struct_cte}
\end{align}
where $h$ is expressed in meters, $v$ is the wind speed in
$\mathrm{m\,s^{-1}}$, and $A\simeq C_n^2(0)$ characterizes the
ground-level turbulence strength. The structure constant has units
$\mathrm{m^{-2/3}}$. For a horizontal channel at approximately constant
altitude, $C_n^2(h)$ can be regarded as constant along the propagation
path.

A common parameter characterizing the strength of turbulence is the
Rytov variance \cite{AndrewsLaserBeam},
\begin{align}
    \sigma_{R,{\rm h}}^2(h,L)
    =
    1.23\,C_n^2(h)\,k^{7/6}L^{11/6},
    \qquad
    k=\frac{2\pi}{\lambda},
    \label{eq:horizontal_rytov}
\end{align}
where $\lambda$ is the optical wavelength. In the following, atmospheric
propagation is restricted to the weak-turbulence regime,
\begin{align}
    \sigma_{R,{\rm h}}^2(h,L)<1.
\end{align}
However we adopt the more conservative threshold $\sigma_{R,h}^2<0.3$ when determining the validity boundaries shown in the numerical results.
\footnote{The classification of turbulence regimes based on Rythov variance is not unique. While $\sigma_R^2<1$ is commonly associated with weak turbulence \cite{AndrewsLaserBeam}, a more restrictive classification distinguishes weak ($\sigma_R^2<0.3$), moderate ($0.3\leq\sigma_R^2\leq1$), and strong ($\sigma_R^2>1$) turbulence \cite{Ji2026Turbulence}.}
A second useful quantity is the spherical-wave coherence length
\cite{fante1975electromagnetic,pirandola2021limits},
\begin{align}
    \rho_0
    =
    \left[
        0.548\,k^2 C_n^2(h)L
    \right]^{-3/5},
    \label{eq:horizontal_rho0}
\end{align}
which characterizes the transverse coherence scale imposed by atmospheric
turbulence. The weak-turbulence propagation model considered below further
requires
\begin{align}
    L
    <
    k\,[\min\{2a_R,\rho_0\}]^2,
    \label{eq:horizontal_weak_condition}
\end{align}
where $a_R$ is the radius of the circular receiver aperture
\cite{fante1975electromagnetic,pirandola2021limits}. Within these
conditions, the beam-broadening and beam-wandering model introduced below
is used to determine the short-term beam width and the statistics of the
beam-centroid displacement.

\begin{remark}
    Throughout this work, we adopt reference atmospheric conditions based on the 1976 US Standard Atmosphere for atmospheric extinction and the Hufnagel--Valley 5/7 profile for turbulence, with the latter restricted to the weak-turbulence regime. We consider aerosol-free conditions and neglect cloud-induced attenuation, without associating the atmospheric profiles with a specific geographical location, season, or time of day. The contribution of ambient optical background radiation is incorporated through the effective dark-count probabilities $p_d^0$ and $p_d^1$, introduced in Sec.~\ref{Sec:numerics}. These parameters account for both intrinsic detector dark counts and background-induced detection events, assuming that the latter can be treated as independent of the optical signals.
\end{remark}

\subsection{Deterministic extinction}

The deterministic factor $\eta_{\ell,h}(h,L)$ in
Eq.~\eqref{eq:horizontal_transmittance} accounts for the attenuation
experienced by the optical signal along the atmospheric propagation path.
We evaluate this contribution using the LOWTRAN7 atmospheric propagation
model \cite{Kneizys1988LOWTRAN}, through the Python implementation from Ref. \cite{PicciaLowtran}. A similar approach has
recently been adopted for the modeling of free-space quantum communication
channels in Ref.~\cite{Karakosta2025}.

LOWTRAN provides the atmospheric transmittance by accounting for the main
extinction mechanisms, including molecular absorption and scattering,
continuum absorption, and aerosol extinction. Their contributions depend
on the optical wavelength, the atmospheric conditions, and the propagation
geometry. We therefore denote the atmospheric extinction as
\begin{align}
    \eta_{\ell,h}(h,L)
    =
    T_{\rm atm}
    \left(
        \lambda;
        h,L,\mathcal{A}
    \right),
    \label{eq:horizontal_extinction}
\end{align}
where $\lambda$ is the optical wavelength, while $\mathcal{A}$ collectively
denotes the atmospheric configuration, including the atmospheric profile
and, when considered, the aerosol model.

Since the transmitter and receiver are assumed to lie approximately at
the same altitude $h$, the atmospheric properties can be regarded as
nearly uniform along the propagation direction. Accordingly,
$T_{\rm atm}$ is evaluated in LOWTRAN using a horizontal
constant-pressure path of length $L$. Under these conditions, the
attenuation can equivalently be expressed through the effective extinction
coefficient
\begin{align}
    \alpha_{\rm LT}(h,\lambda;\mathcal{A})
    :=
    -\frac{1}{L}
    \ln
    T_{\rm atm}
    \left(
        \lambda;
        h,L,\mathcal{A}
    \right),
    \label{eq:lowtran_alpha}
\end{align}
so that
\begin{align}
    \eta_{\ell,h}(h,L)
    =
    \exp\left[
        -\alpha_{\rm LT}(h,\lambda;\mathcal{A})L
    \right].
    \label{eq:lowtran_beer}
\end{align}
Provided that the horizontal atmospheric conditions are sufficiently
uniform for Beer--Lambert scaling to hold, the quantity
$\alpha_{\rm LT}(h,\lambda;\mathcal{A})$ becomes independent of the path
length $L$. The entire dependence on the communication distance is then
contained in the exponential factor, so that a single altitude-dependent
extinction coefficient can be used to evaluate the atmospheric
transmittance for different horizontal link lengths.

At sufficiently high altitudes, the atmospheric contribution becomes
negligible and
\begin{align}
    \eta_{\ell,h}(h,L)\rightarrow 1.
\end{align}
In the atmosphere-free regime, the remaining channel attenuation is determined by diffraction and the fluctuating collection efficiency.

\subsection{Collection efficiency}
The collection-efficiency factor $\eta_{\mathrm{col}}(r)$ in
Eq.~\eqref{eq:horizontal_transmittance} accounts for the fraction of the
propagating optical field captured by the receiver aperture. It depends
both on the beam spot size at the receiver, which is affected by
diffraction and atmospheric turbulence, and on the instantaneous
transverse displacement $r$ of the beam centroid with respect to the
center of the receiver.

We consider a Gaussian beam with initial field spot size $w_0$ and radius
of curvature $R_0$. In the absence of turbulence, diffraction increases
the beam spot size and, after propagation over a distance $L$, one has
\cite{pirandola2021limits}
\begin{align}
    w_L^2
    =
    w_0^2
    \left[
        \left(
            1-\frac{L}{R_0}
        \right)^2
        +
        \left(
            \frac{L}{L_R}
        \right)^2
    \right],
    \qquad
    L_R=\frac{\pi w_0^2}{\lambda},
    \label{eq:horizontal_diffraction}
\end{align}
where $L_R$ is the Rayleigh range. For a collimated beam,
$R_0=\infty$.

In the weak-turbulence regime it is useful to distinguish two physical
effects occurring on different timescales. Small turbulent structures
broaden the beam around its instantaneous position, increasing the
diffraction-limited spot size $w_L$ into the short-term spot size
$w_{\rm st}$. Larger-scale turbulent structures instead induce a random
motion of the optical beam in the transverse plane. Averaging over these
fluctuations leads to the long-term spot size $w_{\rm lt}$.

Following Ref.~\cite{recolons2007beamwander}, we quantify the effective beam-wander contribution to the difference between the long- and short-term beam widths through
\begin{align}
    \Delta_{\rm BW}^2
    :=
    w_{\rm lt}^2-w_{\rm st}^2.
    \label{eq:DeltaBW_definition}
\end{align}
For a generic Gaussian beam, this quantity is modeled as
\begin{align}
    \Delta_{\rm BW}^2
    =
    7.25\,L^2w_0^{-1/3}
    \int_0^L dz\,
    C_n^2(z)
    \left(1-\frac{z}{L}\right)^2
    \left|1-\frac{z}{R_0}\right|^{-1/3}.
    \label{eq:DeltaBW_general}
\end{align}
For the horizontal channel considered here,
$C_n^2(z)=C_n^2(h)$ can be regarded as constant along the propagation
path. Furthermore, for a collimated beam \eqref{eq:DeltaBW_general} reduces to
\begin{align}
    \Delta_{\rm BW}^2
    \simeq
    2.42\,C_n^2(h)L^3w_0^{-1/3}.
    \label{eq:DeltaBW_horizontal}
\end{align}
Within the same weak-fluctuation model, the long-term spot size is
\cite{recolons2007beamwander}
\begin{align}
    w_{\rm lt}^2
    =
    w_L^2
    \left[
        1+
        1.33\,\sigma_{R,{\rm h}}^2
        \Lambda_{\rm B}^{5/6}
    \right],
    \qquad
    \Lambda_{\rm B}
    =
    \frac{2L}{k w_L^2}.
    \label{eq:wlt_recolons}
\end{align}
The short-term spot size entering the collection-efficiency model is
therefore obtained as
\begin{align}
    w_{\rm st}^2
    =
    w_{\rm lt}^2-\Delta_{\rm BW}^2.
    \label{eq:wst_recolons}
\end{align}
The effective beam-wander quantity $\Delta_{\rm BW}^2$ should not be
identified directly with the statistical variance of the beam centroid.
Indeed, Ref.~\cite{recolons2007beamwander} distinguishes between the
displacement of the beam centroid and the motion of the hot spot,
defined as the point of maximum irradiance within the beam profile. The
quantity $\Delta_{\rm BW}^2$ is an effective beam-wander measure
associated with the difference between the long- and short-term beam
widths, and does not coincide with either of these displacement statistics
individually. In the model of Ref.~\cite{recolons2007beamwander}, the
mean-square radial displacement of the centroid is related to it by
$\langle\rho_c^2\rangle\simeq0.56\,\Delta_{\rm BW}^2$. Assuming isotropic
transverse fluctuations, the turbulence-induced variance associated with
either Cartesian component of the centroid displacement is therefore
\begin{align}
    \sigma_{\rm TB}^2
    \simeq
    0.28\,\Delta_{\rm BW}^2.
    \label{eq:sigmaTB}
\end{align}

In addition to turbulence-induced beam wandering, pointing jitter produces
an angular error $\theta_{\rm pe}$. We define $\theta_{\rm pe}$ as the rms
pointing error associated with either Cartesian angular component. Its
contribution to the corresponding transverse-displacement variance at the
receiver is
\begin{align}
    \sigma_{\rm P}^2
    =
    (L\theta_{\rm pe})^2.
    \label{eq:pointing_variance}
\end{align}
Assuming statistically independent turbulence-induced and pointing
fluctuations, the total variance associated with either Cartesian component
of the beam-centroid displacement is
\begin{align}
    \sigma_{\rm w}^2
    =
    \sigma_{\rm TB}^2+\sigma_{\rm P}^2.
    \label{eq:sigma_w}
\end{align}
In the absence of atmospheric turbulence, $C_n^2(h)\rightarrow0$, so that
$\sigma_{R,{\rm h}}^2\rightarrow0$,
$\Delta_{\rm BW}^2\rightarrow0$, and consequently
\begin{align}
    w_{\rm st}
    \rightarrow
    w_{\rm lt}
    \rightarrow
    w_L,
    \qquad
    \sigma_{\rm TB}^2\rightarrow0.
    \label{eq:atmosphere_free_beam}
\end{align}
Pointing fluctuations may nevertheless remain finite. Once $w_{\rm st}$ is known, we define the maximal collection efficiency as
\begin{align}
    \eta_{\mathrm{col}}^{\max}
    =
    1-
    \exp\left(
        -\frac{2a_R^2}{w_{\rm st}^2}
    \right),
    \label{eq:max_collection}
\end{align}
which corresponds to the collection efficiency obtained when the beam
centroid $\vec r_C$ is perfectly aligned with the center $\vec r_R$ of the
circular receiver aperture. In the far-field regime $L\gg L_R$, we define
\begin{align}
    \eta_{\mathrm{col}}^{\mathrm{far}}
    =
    \frac{2a_R^2}{w_{\rm st}^2},
    \label{eq:far_collection}
\end{align}
so that
$\eta_{\mathrm{col}}^{\max}\simeq
\eta_{\mathrm{col}}^{\mathrm{far}}$
when $\eta_{\mathrm{col}}^{\mathrm{far}}\ll1$.

In general, the beam centroid is displaced with respect to the center of
the receiver aperture by
\begin{align}
    r
    =
    \left\|
        \vec r_C-\vec r_R
    \right\|
    \geq0,
    \label{eq:r_displacement}
\end{align}
which implies
$\eta_{\mathrm{col}}(r)\leq\eta_{\mathrm{col}}^{\max}$.
Following Refs.~\cite{vasylyev2012toward,pirandola2021limits}, the
collection efficiency can be approximately expressed as
\begin{align}
    \eta_{\mathrm{col}}(r)
    \simeq
    \eta_{\mathrm{col}}^{\max}
    \exp\left[
        -\left(
            \frac{r}{r_0}
        \right)^\gamma
    \right],
    \label{eq:etacr}
\end{align}
where
\begin{align}
    \gamma
    &=
    \frac{
        4\eta_{\mathrm{col}}^{\mathrm{far}}
        \Lambda_1(\eta_{\mathrm{col}}^{\mathrm{far}})
    }{
        1-\Lambda_0(\eta_{\mathrm{col}}^{\mathrm{far}})
    }
    \left[
        \ln
        \frac{
            2\eta_{\mathrm{col}}^{\max}
        }{
            1-\Lambda_0(\eta_{\mathrm{col}}^{\mathrm{far}})
        }
    \right]^{-1},
    \\
    r_0
    &=
    a_R
    \left[
        \ln
        \frac{
            2\eta_{\mathrm{col}}^{\max}
        }{
            1-\Lambda_0(\eta_{\mathrm{col}}^{\mathrm{far}})
        }
    \right]^{-1/\gamma}.
    \label{eq:gamma_r0}
\end{align}
Here
\begin{align}
    \Lambda_n(x)
    =
    e^{-2x}I_n(2x),
\end{align}
with $I_n$ the modified Bessel function of the first kind of order $n$. The total instantaneous transmittance can therefore be written as
\begin{align}
    \eta_h(r)
    &=
    \eta_h^{\max}
    \exp\left[
        -\left(
            \frac{r}{r_0}
        \right)^\gamma
    \right],
    \label{eq:etatot}
    \\
    \eta_h^{\max}
    &=
    \eta_{\ell,h}(h,L)
    \eta_{\mathrm{col}}^{\max}.
    \label{eq:etamax}
\end{align}
In the atmosphere-free regime, $\eta_{\ell,h}\simeq 1$ and hence $\eta_h^{\max}\simeq\eta_{\mathrm{col}}^{\max}$.

\subsection{PDT in the weak-turbulence regime}

The instantaneous beam-centroid displacement is described by the radial
random variable $R$. Let $(R_x,R_y)$ denote the Cartesian coordinates of
the beam centroid in the transverse receiver plane, measured with respect
to the center of the receiver aperture, so that
\begin{align}
    R^2
    =
    R_x^2+R_y^2.
\end{align}
Assuming zero systematic misalignment and isotropic fluctuations, the two
Cartesian components are modeled as independent Gaussian random variables,
\begin{align}
    R_x
    &\sim
    \mathcal{N}(0,\sigma_{\rm w}^2),
    &
    R_y
    &\sim
    \mathcal{N}(0,\sigma_{\rm w}^2).
    \label{eq:cartesian_displacement}
\end{align}
Here $\sigma_{\rm w}^2$ is the variance of either Cartesian component of
the beam-centroid displacement, as defined in Eq.~\eqref{eq:sigma_w}. The corresponding joint probability density is
\begin{align}
    F_{R_x,R_y}(r_x,r_y)
    =
    \frac{1}{2\pi\sigma_{\rm w}^2}
    \exp\left[
        -\frac{r_x^2+r_y^2}
        {2\sigma_{\rm w}^2}
    \right].
    \label{eq:joint_displacement}
\end{align}
We are interested in the probability density $F_R(r)$ of the radial
displacement $R$. Writing $r_x=r\cos\theta$ and
$r_y=r\sin\theta$ and integrating over the angular degree of freedom gives

\begin{align}
    F_R(r)
    &=
    \int_0^{2\pi}
    F_{R_x,R_y}
    (r\cos\theta,r\sin\theta)\,
    r\,d\theta
    \\
    &=
    \frac{r}{\sigma_{\rm w}^2}
    \exp\left(
        -\frac{r^2}{2\sigma_{\rm w}^2}
    \right),
    \qquad r>0.
    \label{eq:rayleigh_displacement}
\end{align}
This is a Rayleigh distribution, corresponding to a particular case of a
Weibull distribution with shape parameter $k=2$ and scale parameter
$\sqrt{2}\sigma_{\rm w}$.

Let $\mathrm{H}$ denote the random variable associated with the total
transmittance $\eta_h$. To obtain its probability density
$F_{\mathrm{H}}(\eta_h)$, we first invert Eq.~\eqref{eq:etatot}, obtaining
\begin{align}
    r(\eta_h)
    =
    r_0
    \left[
        \ln\left(
            \frac{\eta_h^{\max}}{\eta_h}
        \right)
    \right]^{1/\gamma},
    \qquad
    0<\eta_h\leq\eta_h^{\max}.
    \label{eq:r_of_eta}
\end{align}
A standard change of variables gives
\begin{align}
    F_{\mathrm{H}}(\eta_h)
    =
    F_R\!\left(r(\eta_h)\right)
    \left|
        \frac{dr}{d\eta_h}
    \right|,
    \label{eq:PDT_change_variables}
\end{align}
with
\begin{align}
    \left|
        \frac{dr}{d\eta_h}
    \right|
    =
    \frac{r_0}{\gamma\eta_h}
    \left[
        \ln\left(
            \frac{\eta_h^{\max}}{\eta_h}
        \right)
    \right]^{1/\gamma-1}.
    \label{eq:PDT_jacobian}
\end{align}
We therefore obtain the probability distribution of transmittance
\begin{align}
    F_{\mathrm{H}}(\eta_h)
    &=
    \frac{r_0^2}
    {\sigma_{\rm w}^2\gamma\eta_h}
    \left[
        \ln\left(
            \frac{\eta_h^{\max}}{\eta_h}
        \right)
    \right]^{\frac{2}{\gamma}-1}
    \nonumber\\
    &\quad\times
    \exp\left\{
        -\frac{r_0^2}
        {2\sigma_{\rm w}^2}
        \left[
            \ln\left(
                \frac{\eta_h^{\max}}{\eta_h}
            \right)
        \right]^{\frac{2}{\gamma}}
    \right\},
    \qquad
    0<\eta_h\leq\eta_h^{\max}.
    \label{eq:PDT}
\end{align}
The above probability density is valid for $\sigma_w>0$.
Importantly, Eq.~\eqref{eq:PDT} remains non-trivial in the
atmosphere-free limit. Indeed, when $C_n^2(h)\rightarrow0$ one has
$\rho_0\rightarrow\infty$,
$\Delta_{\rm BW}^2\rightarrow0$,
$w_{\rm st}\rightarrow w_{\rm lt}\rightarrow w_L$, and
$\sigma_{\rm TB}^2\rightarrow0$, whereas
\begin{align}
    \sigma_{\rm w}^2
    \rightarrow
    (L\theta_{\rm pe})^2.
\end{align}
Therefore, a non-zero pointing error is sufficient to produce a fluctuating transmittance even in the atmosphere-free regime. 
The limit case $\theta_{\mathrm{pe}}\rightarrow 0$ must be treated separately. Indeed in the scenario $\sigma_{\mathrm{w}}\rightarrow 0$, implying that the beam remains perfectly centered and the channel becomes a fixed-transmittance channel
\begin{align}
    F_{\mathrm{H}}(\eta_h)
    \longrightarrow
    \delta\left(
        \eta_h-\eta_h^{\max}
    \right).
    \label{eq:deterministic_horizontal_limit}
\end{align}

\subsection{Signal transmittance}
\label{Sec:SignalTransmittance}
The probability distribution derived above characterizes the fluctuations of the channel transmittance. For a fixed realization of the channel with transmittance $\eta$ (dropping subscript $h$ for sake of simplicity), a coherent state $\ket{\gamma}$ is transformed according to $\ket\gamma\longmapsto\ket{\sqrt{\eta}\,\gamma}$. Since the transmittance fluctuates according to the PDT $F_{\mathrm{ H}}(\eta)$, the state at the output of the horizontal channel is obtained by averaging over all the possible realization of the channel:
\begin{align}
    \rho_{\gamma}^{\rm H}
    =
    \int_0^{\eta^{\max}}
    d\eta\,
    F_{\rm H}(\eta)
    \ket{\sqrt{\eta}\,\gamma}
    \bra{\sqrt{\eta}\,\gamma}.
    \label{eq:horizontal_fading_state}
\end{align}
In the MDI configuration considered here, Alice and Bob transmit their states through two horizontal free-space links characterized by the transmittances $\eta_A$ and $\eta_B$, respectively. In general the two links may have different propagation parameters and therefore different transmittance distributions. Assuming statistically independent channel fluctuations, their joint distribution factorizes as 
\begin{align}
    F_{\mathrm{H}_A,\mathrm{H}_B}
    (\eta_A,\eta_B)
    =
    F_{\mathrm{H}_A}(\eta_A)
    F_{\mathrm{H}_B}(\eta_B).
    \label{eq:joint_MDI_PDT}
\end{align}
For coherent states $\ket{\gamma_x}$ and $\ket{\beta_y}$ prepared by Alice and Bob, respectively, the joint state arriving at the relay is therefore
\begin{align}
\rho^{\mathrm H}_{xy}
={}&
\int_0^{\eta_A^{\max}} d\eta_A
\int_0^{\eta_B^{\max}} d\eta_B\,
F_{\mathrm H_A}(\eta_A)
F_{\mathrm H_B}(\eta_B)
\nonumber\\
&\times
\ket{\sqrt{\eta_A}\gamma_x}
\bra{\sqrt{\eta_A}\gamma_x}
\otimes
\ket{\sqrt{\eta_B}\beta_y}
\bra{\sqrt{\eta_B}\beta_y}.
\label{eq:MDI_fading_state}
\end{align}
Let now $\{M^c\}$  denote the POVM describing the measurement performed at the repeater and $U_{\rm BS}$ the beam-splitter transformation preceding the detection. The conditional probability of obtaining the measurement outcome $c$ is then given by
\begin{align}
    \widetilde q(c|x,y)
    =
    \operatorname{Tr}
    \left[
        M^c\,
        U_{\rm BS}
        \rho^{\mathrm H}_{xy}
        U_{\rm BS}^{\dagger}
    \right].
    \label{eq:MDI_fading_probability}
\end{align}
Substituting Eq.~\eqref{eq:MDI_fading_state} into Eq.~\eqref{eq:MDI_fading_probability}, one obtains
\begin{align}
\widetilde q(c|x,y)
={}&
\int_0^{\eta_A^{\max}} d\eta_A
\int_0^{\eta_B^{\max}} d\eta_B\,
F_{\mathrm H_A}(\eta_A)
F_{\mathrm H_B}(\eta_B)
\nonumber\\
&\times
\operatorname{Tr}
\Bigg[
M^c\,
U_{\rm BS}
\Big(
\ket{\sqrt{\eta_A}\gamma_x}
\bra{\sqrt{\eta_A}\gamma_x}
\otimes
\ket{\sqrt{\eta_B}\beta_y}
\bra{\sqrt{\eta_B}\beta_y}
\Big)
U_{\rm BS}^{\dagger}
\Bigg].
\label{eq:MDI_fading_probability_expanded}
\end{align}
For convenience, we define the conditional probability associated with fixed values of the two channel transmittances as
\begin{align}
p(c|x,y,\eta_A,\eta_B)
:=
\operatorname{Tr}
\Bigg[
M^c\,
U_{\rm BS}
\Big(
\ket{\sqrt{\eta_A}\gamma_x}
\bra{\sqrt{\eta_A}\gamma_x}
\otimes
\ket{\sqrt{\eta_B}\beta_y}
\bra{\sqrt{\eta_B}\beta_y}
\Big)
U_{\rm BS}^{\dagger}
\Bigg].
\label{eq:fixed_transmittance_probability}
\end{align}
Hence, the honest conditional probability is obtained by averaging the fixed-transmittance probability over the fluctuations of the two free-space links,
\begin{align}
    \widetilde q(c|x,y)
    =
    \int_0^{\eta_A^{\max}} d\eta_A
    \int_0^{\eta_B^{\max}} d\eta_B\,
    F_{\mathrm H_A}(\eta_A)
    F_{\mathrm H_B}(\eta_B)\,
    p(c|x,y,\eta_A,\eta_B).
    \label{eq:MDI_fading_probability_average}
\end{align}
Equation~\eqref{eq:MDI_fading_probability_average} provides the connection between the horizontal-channel model and the honest probabilities entering the MDI QKD protocol. The explicit form of $p(c|x,y,\eta_A,\eta_B)$ depends on the measurement operators adopted for the relay and will be specified in Sec. \ref{Sec:numerics}.

\section{Numerical framework}\label{Sec:numerics}
Provided the framework for the variable-length analysis, we can now showcase our approach by calculating secret-key rates according to \eqref{eq:expected_SKR} and our conic formulation \cite{Lorente2025quantumkey,Navarro2025}. We set the security coefficients according to their optimal values \cite{goh2026simpleqkdkeyratecomputations}

\begin{align}
    \epsilon_\mathrm{EC} = \frac{\alpha-1}{2 \alpha-1} \epsilon  &&
    \epsilon_\mathrm{PA} = \frac{\alpha}{2 \alpha-1} \epsilon
\end{align}
which together define an $\epsilon-$secure implementation of the protocol. Following standard practices for QKD, we set $\epsilon=10^{-10}$.

Now, specifying the key maps entering the FastRényiQKD cone, we include local randomization on Alice's raw-key variable, with bit-flip probability $p_{\mathrm{LR}}\in[0, 1/2]$, for an evaluation of its impact on the secret key rate \cite{staffieri2026}. The complete construction is reported in Appendix \ref{App:NoisyPreproc}, while here we only introduce the maps needed for the numerical implementation. Let
\begin{align}
\Lambda_{0}^{(p_{\mathrm{LR}})}
&=
(1-p_{\mathrm{LR}})Z^0+p_{\mathrm{LR}}Z^1,
&
\Lambda_{1}^{(p_{\mathrm{LR}})}
&=
p_{\mathrm{LR}}Z^0+(1-p_{\mathrm{LR}})Z^1,
\end{align}
where the operators $Z^k, \, k=0,1$, select Alice and Bob's two key sectors. Introducing the two dimensional register $R$ (see Section \ref{sec:ProtDescription}), we define 
\begin{align}
K^{(p_{\mathrm{LR}})}
:=
\sum_{r=0,1}
\ket{r}_{R}
\otimes
\sqrt{\Lambda_{r}^{(p_{\mathrm{LR}})}}.
\end{align}
The key map is then 
\begin{align}
\mathcal G^{(p_{\mathrm{LR}})}(\cdot)
=
K^{(p_{\mathrm{LR}})}(\cdot)K^{(p_{\mathrm{LR}})\dagger}.
\label{eq:G-noisy-main}
\end{align}
Here we treat the two relay announcements $c'=0,1$ symmetrically, and therefore $\mathcal G^{0,(p_{\mathrm{LR}})}=\mathcal G^{1,(p_{\mathrm{LR}})}=\mathcal G^{(p_{\mathrm{LR}})}$. The corresponding pinched map is 
\begin{align}
\left(
\mathcal Z
\circ
\mathcal G^{(p_{\mathrm{LR}})}
\right)(\cdot)
=
\sum_{r=0,1}
\pro{r}_{R}
\otimes
\sqrt{\Lambda_{r}^{(p_{\mathrm{LR}})}}
(\cdot)
\sqrt{\Lambda_{r}^{(p_{\mathrm{LR}})}}.
\label{eq:ZG-noisy-main}
\end{align}
However, these maps have a nullspace, which means that strict feasibility is not guaranteed  for the optimization and therefore convergence cannot be achieved via interior point methods. We overcome this limitation by performing facial reduction \cite{hu2022,Lorente2025quantumkey}. Denoting by $W_{G,p_{\mathrm{LR}}}$ and $W_{ZG,p_{\mathrm{LR}}}$ the facial reduction isometries associated  respectively to $
\mathcal G^{(p_{\mathrm{LR}})}$ and $\mathcal Z^{(p_{\mathrm{LR}})}
\circ
\mathcal G^{(p_{\mathrm{LR}})}$ we define 
\begin{align}
    \widehat S_{p_{\mathrm{LR}}}
:=
W_{ZG,p_{\mathrm{LR}}}^{\dagger}
W_{G,p_{\mathrm{LR}}}.
\label{eq:S-main}
\end{align}
The explicit construction of both isometries, together with the determination of the corresponding output supports is provided in Appendix \ref{App:NoisyPreproc}. The facially reduced maps entering the conic optimization are obtained
by compressing the original maps onto their effective output spaces,
namely
\begin{align}
\gmap^{c'}(\cdot)
&:=
W_{G,p_{\mathrm{LR}}}^{\dagger}\,
\mathcal G^{(p_{\mathrm{LR}})}(\cdot)
W_{G,p_{\mathrm{LR}}},
\label{eq:Ghat-main}
\\
\zgmap^{c'}(\cdot)
&:=
W_{ZG,p_{\mathrm{LR}}}^{\dagger}
\left(
\mathcal Z
\circ
\mathcal G^{(p_{\mathrm{LR}})}
\right)(\cdot)
W_{ZG,p_{\mathrm{LR}}},
\label{eq:ZGhat-main}
\end{align}
for $c'=0,1$. Next, we consider set of reference states for each honest party. Following \cite[Appendix D]{Pereira2020correlated} we set

\begin{subequations}
\begin{align}
    \ket{\phi_0} &= \ket{\gamma} \\
    \ket{\phi_1} &= \ket{-\gamma e^{i \delta}} \\
    \ket{\phi_2} &= \ket{\oslash},
\end{align}
\end{subequations}
where $\delta\in [0,2\pi)$, represents a possible phase deviation, caused by an imperfect preparation at the source.
Each joint reference $\ket{\phi_{xy}}$ is thereby built via a tensor product \eqref{eq:jointReference}, and provided by the simplified tolerance margins
\begin{subequations}
    \begin{align}
    \varepsilon_{xy} &= 2\varepsilon, \quad  x,y \in \{0,1\} \label{eq:2e} \\
     \varepsilon_{xy} &= \varepsilon, \quad \; \; x=2, y\neq 2 \text{ or } y = 2, x\neq 2 \label{eq:1e} \\
     \varepsilon_{xy} &= 0  \qquad    x=y=2 \label{eq:0e}
    \end{align}
\end{subequations}
which complete the partial source description.

Given the structure of \eqref{eq:simplifyepsilons}, when the vacuum is chosen (i.e. $p_\mathrm{v}\neq0$), the reference states form a linearly independent set allowing $G$ to being full-rank, which is a crucial feature as a linear dependency induces a singular $G$, such that further facial reduction would be needed for the Gram matrix \cite[Appendix B] {pereira2025optimalkeyratesquantum}. However,  certain scenarios yield $p_\mathrm{v}=0$ as the optimal vacuum probability, such that the reference states form a linearly dependent set.  Nevertheless, owing numerical noise, the conic solver was able to reach a feasible solution even in absence of facial reduction for the Gram matrix,   only implying a penalty on the final key of around $0.1\%$.

To conclude with the numerical framework, we explicitly construct the honest reference
probability distribution $q_{\hat C}$ introduced in
Section~\ref{sec:FiniteSize}. We split its components
into key-generation and testing events as follows:
\begin{align}
    \{q_{\hat C}(\hat c)\}_{\hat c\in\hat{\mc C}}
    =
    \left(
    \{\pkey\tilde q(\bot,\bot,c')\}_{c'=0,1},
    \{\hat p(x,y,c)\tilde q(x,y,c)\}_{(x,y,c)\in\tilde{\mc C}}
    \right).
\end{align}
where we use the masking factor
\begin{align}
    \hat{p}(x,y,c) &= \begin{cases}
        1 - \pkey, &  (x,y,c) \in \{0,1\}^3   \\
        1, & \mathrm{else}
    \end{cases}
\end{align}
In the case of key rounds, the probabilities can be determined by using the relation
\begin{align}
    \tilde{q}(\bot,\bot,c') = \sum_{x,y=0,1} \tilde{q} (x,y,c'), \quad c' = 0,1.
\end{align}
For genuine testing rounds, we may decompose the probabilities according to the independence of Alice and Bob's preparations via the conditioning 
\begin{align}
    \tilde{q}(x,y,c) &= \tilde{q}(c|x,y) \tilde{q}(x) \tilde{q}(y), \quad (x,y,c)\in \hat{\mc{C}}.
\end{align}
Here we directly split the preparations of Alice and Bob, since these are performed independently, and are provided by the set $((1 - p_\mathrm{v})/2,(1- p_\mathrm{v})/2,p_\mathrm{v})$.

Regarding the remaining probabilities, let us provide their explicit values by considering the trusted dark count probabilities $p^0_d,p^1_d \in [0,1]$ for the two detectors at the relay. Hence, the initial POVM for the relay $\{P^\mathrm{nc}, P^0, P^1, P^\mathrm{dc}\}$ becomes, according to the assignations of Alice and Bob for the announcements 

\begin{align}
\begin{pmatrix}
M^\perp\\
M^0\\
M^1
\end{pmatrix}
=
\begin{pmatrix}
1-(1-p_d^0)p_d^1 - (1-p_d^1)p_d^0
&
p_d^1
&
p_d^0
&
1
\\
p_d^0(1-p_d^1)
&
1-p_d^1
&
0
&
0
\\
(1-p_d^0)p_d^1
&
0
&
1-p_d^0
&
0
\end{pmatrix}
\begin{pmatrix}
P^{\rm nc}\\
P^0\\
P^1\\
P^{\rm dc}
\end{pmatrix}.
\label{eq:POVM_transf}
\end{align}
Provided the POVM, and our model for the free-space quantum channel (as provided in Section \ref{sec:Horizontal}), we find the probabilities

\begin{align}
\widetilde q(0|x,y)
=&\,
(1-p_d^1)\,
\mathcal J_1(x,y)
-
(1-p_d^0)(1-p_d^1)\,
\mathcal J_{\rm nc}(x,y),
\label{eq:q0_integrated}
\\
\widetilde q(1|x,y)
=&\,
(1-p_d^0)\,
\mathcal J_0(x,y)
-
(1-p_d^0)(1-p_d^1)\,
\mathcal J_{\rm nc}(x,y),
\label{eq:q1_integrated}
\\
\widetilde q(\perp|x,y)
=&\,
1
-
(1-p_d^0)\,
\mathcal J_0(x,y)
-
(1-p_d^1)\,
\mathcal J_1(x,y)
\nonumber\\
&+
2(1-p_d^0)(1-p_d^1)\,
\mathcal J_{\rm nc}(x,y),
\label{eq:qperp_integrated}
\end{align}
where the quantities $\mathcal J_c(x,y)$, $c\in \{0,1,\mathrm{nc} \}$, are defined and discussed in Appendix~\ref{app:q_honest}, together with the remaining details of the calculation.

\section{Results}
\label{Sec:Results}

We organize our numerical results into two main configurations, referred to as symmetric and asymmetric, according to the communication distances and the signal amplitudes of Alice and Bob. The symmetric configuration is defined by equal distances from the relay, $L_A = L_B$, and equal coherent state amplitude $\gamma_A = \gamma_B$. Conversely, the asymmetric configurations includes any deviation from these conditions, \emph{i.e.} different communication distances, different signal amplitudes or both. Throughout these analyses, we consider the nominal phase deviation as $\delta =0$. Nevertheless source imperfections are still accounted through the fidelity tolerances introduces in the Subsection \ref{Subsec:sourceImperfections}.

We first investigate the symmetric configuration through two complementary analyses of the secret-key rate (SKR) as a function of the communication distance. In the first, we fix the communication altitude and vary the total number of the protocol rounds $n$ assessing the finite-size dependence of the achievable SKR. In the second, we fix $n$ and consider different link altitudes, investigating the influence of the horizontal channel conditions on the protocol performances. As a third, separate study we examine how varying the initial beam waist $w_0$ and the receiver aperture radius $a_R$ affects the achievable SKR and the maximum communication range, considering different optical terminal configurations.

We turn then to the asymmetric configuration, considering all eight possible combinations in which Alice's communications distance and state signal amplitude can be greater than, equal to or smaller than Bob's. We initially compare the SKR as a function of communication distance for all these configurations at fixed altitude and block size $n$, identifying the best performing setup. We examine this configurations at fixed  $n$ for different communications altitudes. Finally, we investigate how different degrees of asymmetry in the communication distances and signal amplitudes affect the achievable SKR.

\subsection{Outline of parameters}

The parameters that are fixed for the protocol  and the horizontal-channel are summarized in Table~\ref{tab:simulation_parameters}. Atmospheric extinction is evaluated using
LOWTRAN7~\cite{Kneizys1988LOWTRAN}, through
the Python implementation of~\cite{PicciaLowtran}.
We adopt the 1976 US Standard Atmosphere,
considering aerosol-free conditions and
horizontal constant-pressure propagation.
The atmospheric extinction is evaluated
at the operating wavelength of $\lambda = 800\,\mathrm{nm}$.
Atmospheric turbulence is described by the
Hufnagel--Valley model, with the parameters
reported in Table~\ref{tab:simulation_parameters}.
Unless explicitly stated otherwise, the
parameters $\{\pkey, p_{\mathrm{v}},\,p_{\mathrm{LR}},\,\alpha,\, \gamma_A,\,\gamma_B\}$
are optimized subject to the constraints
of each configuration.

\begin{remark}
    In all simulations, the dark count probability is fixed to $p_d^0=p_d^1=10^{-5}$, independently of the communication altitude and distance. This choice provides a common low-background reference scenario, allowing us to isolate the effects of atmospheric attenuation, turbulence, and channel asymmetry on the achievable SKR. Consequently, the results should be interpreted as comparative performance predictions under fixed noise conditions, rather than as estimates for daytime or nighttime operation at a particular height.
\end{remark}

\begin{table}[t]
\centering
\begin{tabular}{@{}lll@{}}
\toprule
Parameter & Value & Description \\
\midrule
\multicolumn{3}{c}{\textit{Protocol parameters}} \\
\midrule
$p_d^0=p_d^1$ & $10^{-5}$ &
Detector dark-count probabilities \\
$\varepsilon$ & $10^{-5}$ &
Source-characterization tolerance \\
$\epsilon$
& $10^{-10}$ & Security parameter \\
$f_\mathrm{EC}$ & $1.16$ & Reconciliation inefficiency \\
$\delta$ & $0$ & Nominal phase deviation \\
\midrule
\multicolumn{3}{c}{\textit{Horizontal-channel parameters}} \\
\midrule
$\lambda$ & $800\,\mathrm{nm}$ &
Optical wavelength \\
$w_0$ & $5\,\mathrm{cm}$ &
Initial beam spot radius \\
$R_0$ & $\infty$ &
Initial wavefront radius of curvature \\
$a_R$ & $5\,\mathrm{cm}$ &
Receiver aperture radius \\
$\theta_{\mathrm{pe}}$ & $1\,\mu\mathrm{rad}$ &
RMS pointing error per angular component \\
$\mathcal A$ & US Standard 1976 &
LOWTRAN atmospheric profile \\
$v$ & $21\,\mathrm{m/s}$ &
Hufnagel--Valley wind speed \\
$A$ & $1.7\times10^{-14}\,\mathrm{m}^{-2/3}$ &
Hufnagel--Valley ground-level parameter \\
\bottomrule
\end{tabular}
\caption{Fixed protocol and horizontal-channel
parameters used in the numerical simulations.}
\label{tab:simulation_parameters}
\end{table}

\subsection{Symmetric MDI setup}
\label{subsec:symm}
\begin{figure}[!h]
    \centering
\includegraphics[width=1.0\linewidth]{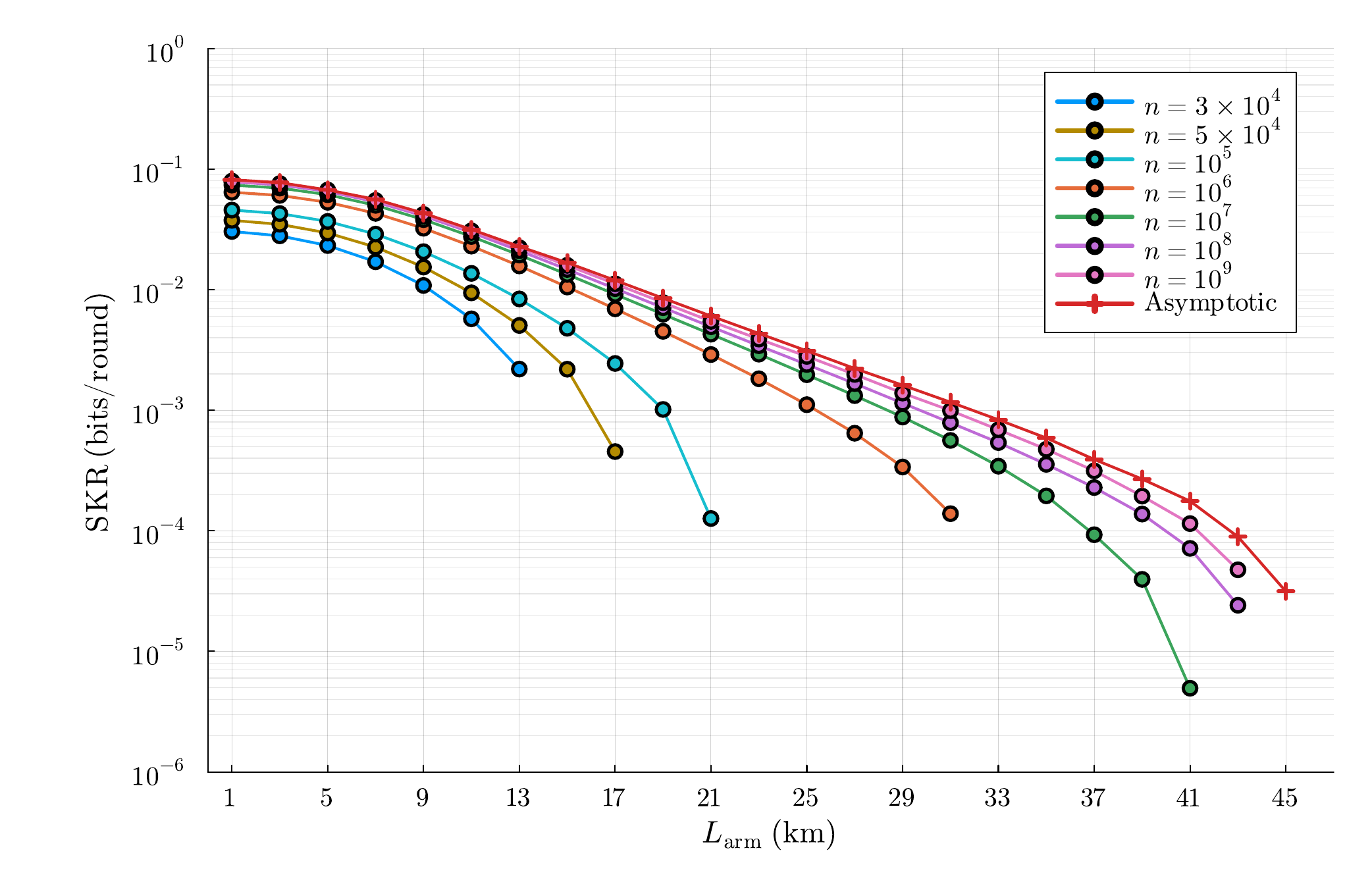}
    \caption{Secret-key rate (SKR) as a function of the individual arm length $L_{\mathrm{arm}}$ for the symmetric MDI QKD configuration ($L_A=L_B=L_{\mathrm{arm}}$ and $\gamma_A=\gamma_B$), considering different block sizes $n$. The simulations are performed at $h=30\,\mathrm{km}$, assuming an atmosphere-free channel while retaining pointing fluctuations. The protocol parameters $\pkey$, $p_{\mathrm{v}}$, $\alpha$, $p_{\mathrm{LR}}$ and the coherent-state amplitude $\gamma_A=\gamma_B$ are optimized at each point. The remaining parameters are reported in Table~\ref{tab:simulation_parameters}.}
\label{fig:FirstFigure}
\end{figure}

\begin{figure}[!h]
    \centering
    \includegraphics[width=1.0\linewidth]{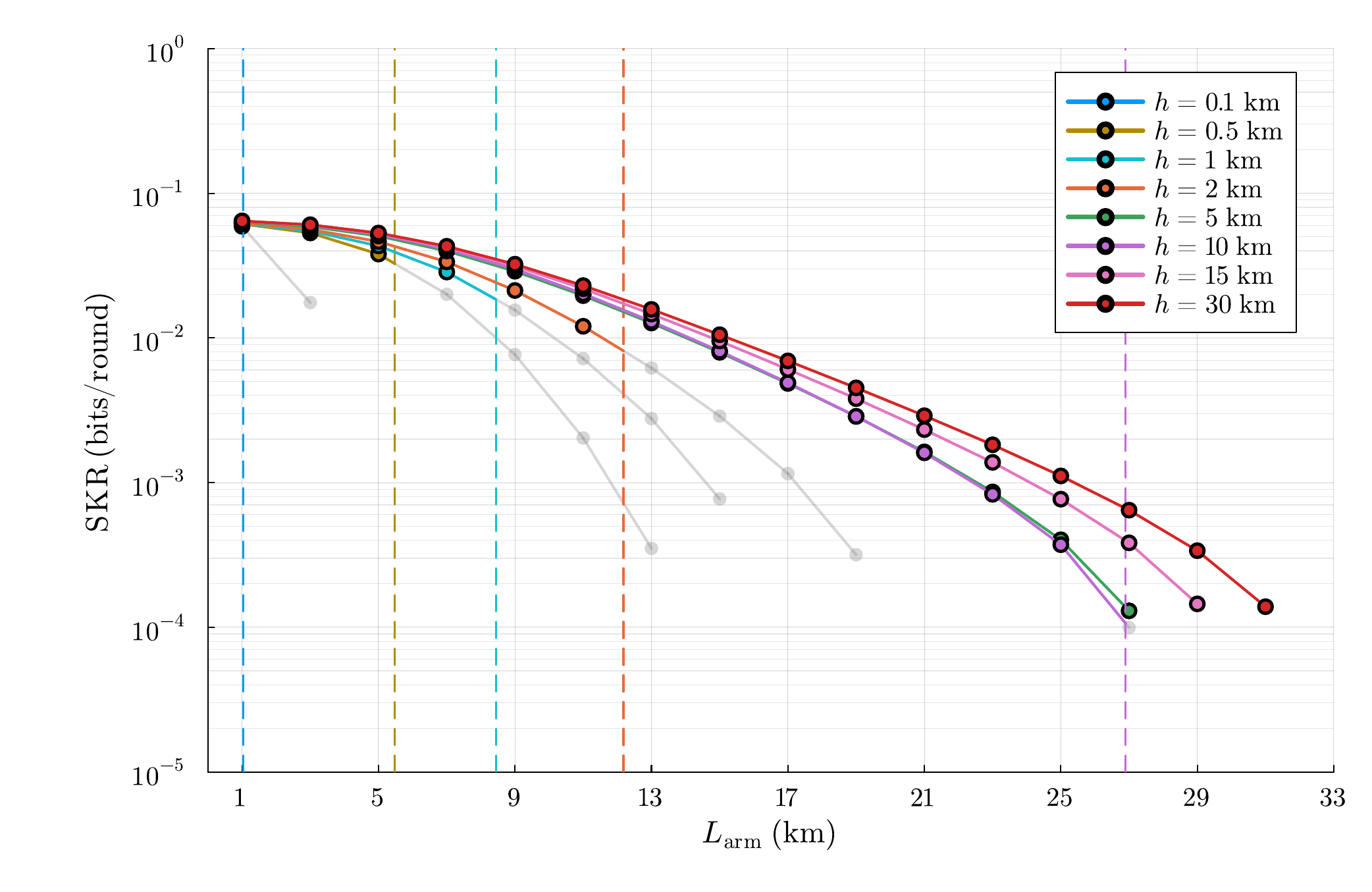}
    \caption{Secret-key rate (SKR) as a function of the
    individual arm length $L$ for the symmetric MDI QKD
    configuration at different propagation altitudes $h$, with a fixed block size $n=10^6$. Vertical dashed lines indicate the maximum arm lengths compatible with the validity conditions of the weak-turbulence model. Points beyond these limits are shown in grey and are excluded from the physically valid regime. The coherent-state amplitude are optimized at each operating point subject to the constrain $\gamma_A=\gamma_B$. The remaining parameters are reported in Table~\ref{tab:simulation_parameters}.}
    \label{fig:SecondFigure}
\end{figure}

For the symmetric configuration we analyze  the trend of the SKR within the communication distance $L$. Fig. \ref{fig:FirstFigure} shows the finite-size dependence of the SKR for horizontal links at a fixed height of $h=30 
\text{km}$. At this altitude, we adopt an atmosphere-free
approximation, motivated by the negligible
atmospheric contribution at high altitudes
discussed in ~\cite[Sec.~II.C]
{Pirandola2021Satellite} 
\footnote{Our numerical simulations further support
this assumption. Indeed, we omit the curve at
$h=20\,\mathrm{km}$ from Fig.~\ref{fig:SecondFigure},
as its SKR nearly overlaps with that obtained
at $h=30\,\mathrm{km}$ in the atmosphere-free regime.}. Accordingly, we assume ideal atmospheric extinction $\eta_{\ell,h=30\text{km}}=1$ and
turbulence effects are neglected as $C_n^2(h=30\text{km})=0$. The SKR is plotted as a function of the individual arm length $L_{\mathrm{arm}}$, so that the total Alice--Bob communication distance is actually $L=2L_{\mathrm{arm}}$. 

As provided in Fig.~\ref{fig:FirstFigure},
increasing the block size improves the
achievable SKR at a given distance and
extends the maximum communication range
over which a positive secret key can be
extracted. This behavior reflects the
reduction of finite-size penalties as
the number of protocol rounds increases. Importantly, simulations achieve positive keys even for a block size $n=3\times10^4$, in line with reduced block sizes obtained in Ref. \cite{kamin25MEATsecurity}, were positive SKRs are achieved for $n=10^5$. Observing SKRs at small block-sizes is remarkable for free space communications, particularly for links involving moving platforms with limited communication windows. Finally, the SKR exhibits a tendency towards saturation beyond $n=10^9$, suggesting that finite-size corrections become relatively negligible for larger block sizes.

We next investigate the influence of
atmospheric conditions by fixing the
block size to $n=10^6$ and evaluating
the SKR at different communication
altitudes. Fig.~\ref{fig:SecondFigure} shows that
atmospheric conditions significantly affect
both the achievable SKR and the maximum
communication range. In general, propagation
at lower altitudes leads to stronger
atmospheric extinction and turbulence,
resulting in a faster degradation of the SKR
with distance. For atmospheric horizontal links, the vertical dashed
lines indicate the maximum arm lengths
compatible with the weak-turbulence conditions
introduced in
Eqs.~\eqref{eq:horizontal_rytov}
and~\eqref{eq:horizontal_weak_condition}.
Points outside the validity regime (i.e. generating positive keys without meeting the weak turbulence assumption) are
displayed in grey and are not considered
physically reasonable predictions of the
adopted channel model.

Interestingly, the relative ordering of the SKR
curves at $h=5\,\mathrm{km}$ and
$h=10\,\mathrm{km}$ reverses at longer distances.
This behavior can be explained by the
non-monotonic altitude dependence of the
Hufnagel--Valley turbulence profile in Eq.~\eqref{eq_ref_index_struct_cte}.
In particular, the model accounts for an
enhancement of atmospheric turbulence near
the tropopause, around $10\,\mathrm{km}$,
associated with high-altitude winds
\cite{CantosRoman2026Atmospheric}.
For the parameters used in our simulations,
$C_n^2(h=5\,\mathrm{km})\simeq
1.20\times10^{-17}\,\mathrm{m}^{-2/3}$,
whereas
$C_n^2(h=10\,\mathrm{km})\simeq
1.67\times10^{-17}\,\mathrm{m}^{-2/3}$.
Consequently, although propagation at
$10\,\mathrm{km}$ benefits from lower
atmospheric extinction, it is subject to
stronger turbulence, resulting in competing
effects on the achievable SKR.

\begin{figure}[!h]
    \centering
    \includegraphics[width=1.0\linewidth]{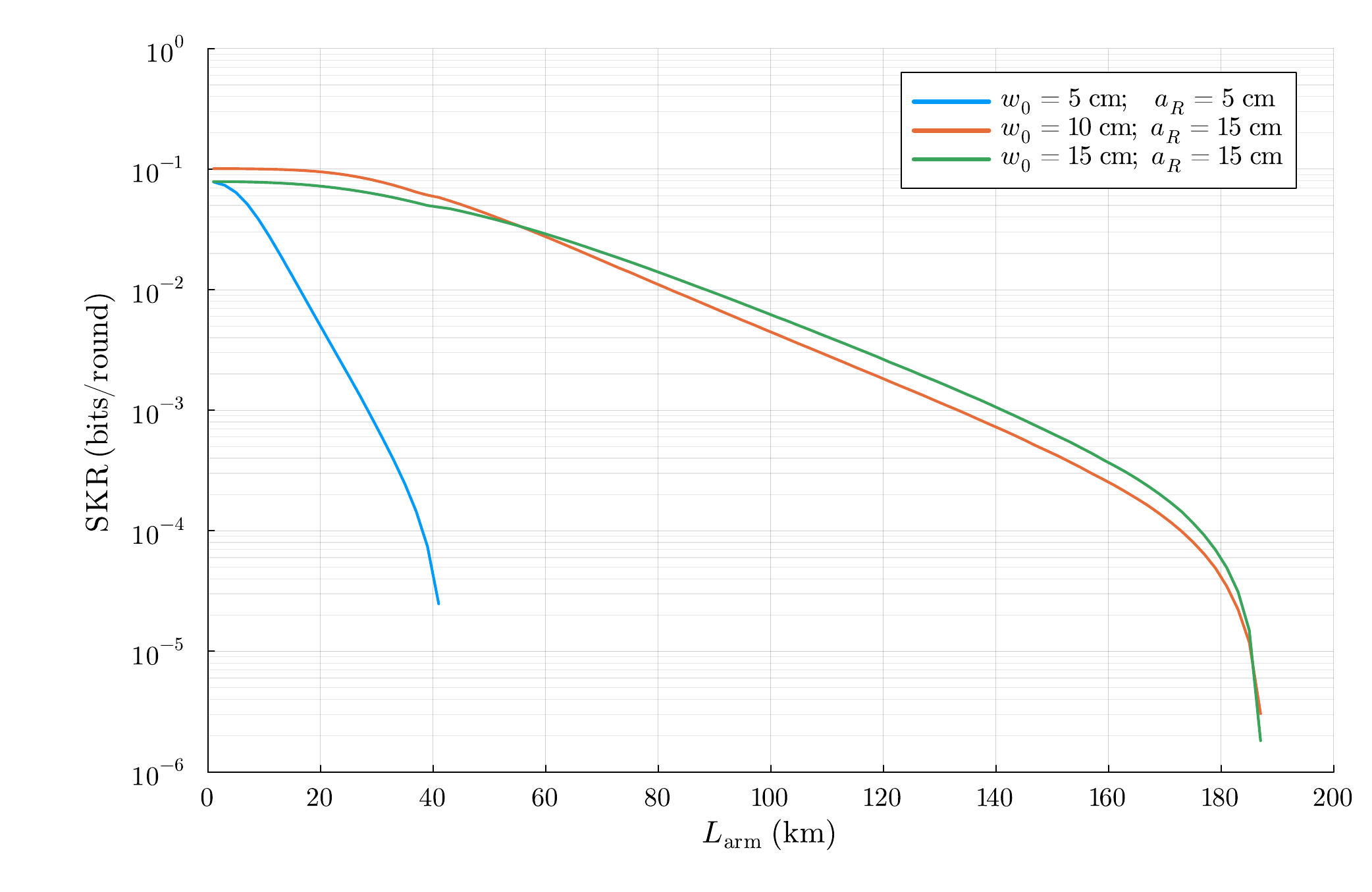}
    \caption{Secret-key rate (SKR) as a function of the individual arm length $L_{\mathrm{arm}}$ for the symmetric MDI QKD configuration, considering different values for the initial beam spot radius $w_0$ and the receiver aperture radius $a_R$. The simulations are performed at $h=30\,\mathrm{km}$, assuming an atmosphere-free channel, with block size $n=10^8$ and optical wavelength $\lambda=850 \; \mathrm{nm}$. The remaining parameters are reported in Table~\ref{tab:simulation_parameters}.}
\label{fig:Plot_6}
\end{figure}

We note that the communication distances achieved with our baseline parameters are relatively short compared with conventional inter-satellite optical links, which commonly span distances of the order of $10^3$ km \cite{Hasler2022LCT}. The achievable communication range depends strongly on the initial beam waist $w_0$ and receiver aperture radius $a_R$. The baseline values of these parameters, reported in Table~\ref{tab:simulation_parameters}, follow horizontal links specifications in \cite[Sec.~V]{gonzalez2024satellite} in line with small satellites \cite{Neumann2018Q3Sat,Roubal2026QUBEII}. Although we retain these values in the subsequent simulations of asymmetric configurations, Fig.~\ref{fig:Plot_6} shows that larger communication distances can be achieved by increasing the optical dimensions, consistently with the larger apertures considered in other quantum communications studies with aerial platforms \cite{Karakosta2025}. This analysis also illustrates how the performance depends on the optical terminal specifications, which vary considerably across satellite systems, including compact CubeSat designs \cite{Trinh2022CubeSOTA,Serra2021CubeSat,Mendes2024CubeSat}. 

Setting the wavelength to $\lambda=850$ nm in Fig.~\ref{fig:Plot_6}, we observe that larger values of the receiver aperture $a_R$ and the initial beam waist $w_0$ allow longer communication ranges within the parameter range investigated. At shorter distances, a smaller $w_0$ yields higher SKRs, until the curves cross at approximately $55\,\mathrm{km}$, beyond which a larger beam waist becomes advantageous. This behavior results from the interplay between the initial beam width and diffraction. A smaller $w_0$ produces a more concentrated beam at short distances, improving the collection efficiency at the receiver. However, it also increases the beam divergence, leading to greater beam spreading and reduced collection efficiency at longer distances. This explains the observed change in the relative SKR performance under the considered atmosphere-free conditions. 

Finally, we point out that simulations for the symmetric configuration yield $p_{\mathrm{LR}}=0$ as the optimal value for the local randomization probability. Hence randomization procedure is not beneficial for the SKR for symmetric MDI setups. 
\footnote{Since we employ direct reconciliation, Alice's raw key defines the secret key. Applying additional independent randomization to Bob's raw key cannot reduce Eve's information about Alice's key, while it can only decrease the correlations between Alice and Bob. Therefore, local randomization on Bob's side cannot improve the SKR.}

\subsection{Asymmetric MDI setup}
We now investigate the impact of asymmetric configurations on the achievable SKR. We consider two degrees of asymmetry: the distances $L_A$ and $L_B$ separating Alice and Bob from the relay, and their transmitted coherent-state amplitudes $\gamma_A$ and $\gamma_B$. Since the nominal phase deviation is fixed to $\delta=0$, the analysis focuses exclusively on these two contributions. We express the communication range in terms of the total Alice--Bob distance, $L=L_A+L_B$, and investigate eight possible asymmetric configurations, together with the symmetric reference case.
\begin{figure}[!h]
    \centering
    \includegraphics[width=1.0\linewidth]{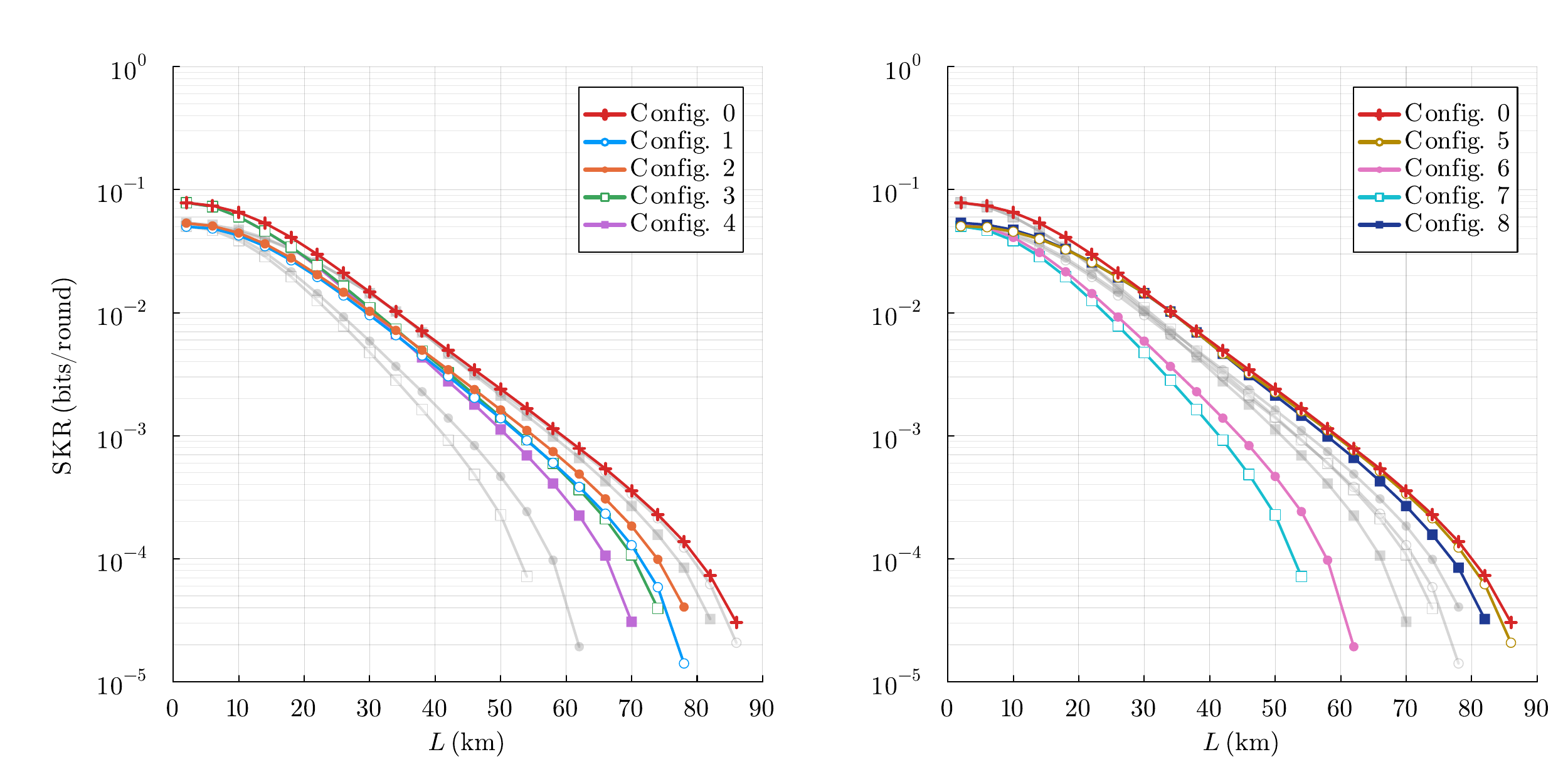}
    \caption{Secret-key rate (SKR) as a function of the total Alice--Bob distance $L=L_A+L_B$ for eight asymmetric MDI QKD configurations, compared with the symmetric reference ($L_A=L_B$ and $\gamma_A=\gamma_B$). The left panel shows configurations with a single degree of asymmetry, either in the arm lengths or in the transmitted amplitudes, while the right panel shows configurations combining both degrees of asymmetry. We consider the arm-length ratios $L_A/L_B\in\{\tfrac{2}{3},1,\tfrac{3}{2}\}$ and the amplitude ratios $\gamma_A/\gamma_B\in\{\tfrac{4}{5},1,\tfrac{5}{4}\}$. Simulations are performed at $h=30\,\mathrm{km}$ in the atmosphere-free regime, with a block size $n=10^8$. The coherent-state amplitudes are optimized at each operating point, subject to the specified ratios. The remaining simulation parameters are reported in Table~\ref{tab:simulation_parameters}.}
    \label{fig:8_confi_asymmetry}
\end{figure}

To interpret their relative performance, we consider the mean photon number in the nominally dark output port of the relay. As derived in Appendix~\ref{app:q_honest}, the beam-splitter output intensities are $\mu_0=|\zeta|^2$ and $\mu_1=|\xi|^2$. For signal--signal preparations with $x=y$, constructive interference occurs at detector $D_0$, so that $\mu_{\mathrm{dark}}=\mu_1$. Conversely, for $x\neq y$, the output ports exchange roles and $\mu_{\mathrm{dark}}=\mu_0$. In both cases, the mean photon number in the nominally dark port is $\langle \mu_{\mathrm{dark}}\rangle = \frac12\left[
\gamma_A^2\langle\eta_A\rangle
+\gamma_B^2\langle\eta_B\rangle
-2\gamma_A\gamma_B
\left\langle\sqrt{\eta_A\eta_B}\right\rangle
\right]$. Here, the average is taken over the fluctuating transmittances of the two channels. This quantity characterises the residual optical intensity arising from imperfect destructive interference. In particular, the residual intensity is generally reduced when the received optical fields are balanced on average, as in the symmetric configuration, whereas differences in the transmitted amplitudes or channel attenuations can increase it and, consequently, the probability of erroneous detector clicks.
For this reason, we are motivated to investigate all possible kind of misalignment between the received signal amplitude at the relay, allowing all the possible configurations that can stem from $\gamma_A\lesseqgtr \gamma B$ and $L_A \lesseqgtr L_B$.

Fig. \ref{fig:8_confi_asymmetry} compares the SKR of the eight asymmetric configurations listed on Table \ref{table:asymm_config} , with the symmetric reference. 
\begin{table}[h!]
\centering
\footnotesize
\begin{tabular}{c c c}
\toprule
Configuration & Arm-length relation & Amplitude relation\\
\midrule
0 (Symmetric) & $L_A=L_B$ & $\gamma_A=\gamma_B$ \\
1 & $L_A=L_B$ & $\gamma_A>\gamma_B$ \\
2 & $L_A=L_B$ & $\gamma_A<\gamma_B$ \\
3 & $L_A>L_B$ & $\gamma_A=\gamma_B$ \\
4 & $L_A<L_B$ & $\gamma_A=\gamma_B$ \\
5 & $L_A>L_B$ & $\gamma_A>\gamma_B$ \\
6 & $L_A>L_B$ & $\gamma_A<\gamma_B$ \\
7 & $L_A<L_B$ & $\gamma_A>\gamma_B$ \\
8 & $L_A<L_B$ & $\gamma_A<\gamma_B$ \\
\bottomrule
\end{tabular}%
\caption{Arm-length and amplitude relations between Alice's and Bob's sides of an asymmetric MDI setup for the 8 possible configurations of asymmetry.}
\label{table:asymm_config}
\end{table}
Specifically, we consider the arm-length ratios $L_A/L_B\in\{\tfrac{2}{3},1,\tfrac{3}{2}\}$ and the amplitude ratios $\gamma_A/\gamma_B\in\{\tfrac{4}{5},1,\tfrac{5}{4}\}$. Results are obtained for $n=10^8$ at $h=30\text{km}$, in the atmosphere-free regime. At short communication distances, configurations with equal transmitted amplitudes exhibit rates close to the symmetric case, since the difference in attenuation between the two arms remains relatively small (see configurations 3 and 4 on the left panel) . Introducing different amplitudes generally produces a stronger imbalance between the fields reaching the relay and hence a larger $\langle\mu_{\mathrm{dark}}\rangle$, resulting in a reduction of the SKR. However, as the communication distance increases, the relative attenuation between unequal arms becomes more pronounced, and the ordering of the curves changes. 

In this regime, an appropriate asymmetry in the transmitted amplitudes can partially compensate for the difference in channel transmittances. In particular, the configurations 5 and 8 provide improved interference conditions by assigning the larger transmitted amplitude to the longer, more attenuated link. This reduces the residual dark-port intensity and allows these configurations to achieve higher SKRs, approaching the performance of the symmetric reference. Conversely, when the smaller amplitude is transmitted through the longer arm, the two sources of asymmetry reinforce each other, increasing $\langle\mu_{\mathrm{dark}}\rangle$ and producing the lowest rates among the configurations considered (see configurations 6 and 7 on the right panel).

The simulations also show that local randomization can improve the SKR in several asymmetric configurations. Unlike the symmetric case, where the optimised preprocessing probability is generally zero, a non-vanishing $p_{\mathrm{LR}}$ can become advantageous when asymmetry modifies the balance between the certified entropy and the error-correction leakage. Table~\ref{tab:qbf_comparison} summarises the corresponding improvements obtained by comparing the optimised results with simulations performed at $p_{\mathrm{LR}}=0$.

\begin{table}[ht]
\centering
\resizebox{\textwidth}{!}{%
\begin{tabular}{c c c c c c c c}
\toprule
Config. & Activation $L$ (km) & Points with active $p_{LR}$ & Mean $p_{LR}$ (active) & Max.\ $p_{LR}$ & Max.\ gain & Average SKR gain (only in active region) \\
\midrule
0 & 86 & 4.5\,\% (1/22)  & 0.0020 & 0.002 & +3.50\,\%  & +3.50\,\% \\
1 & 50 & 40.0\,\% (8/20) & 0.0028 & 0.006 & +14.78\,\% & +0.18\,\% \\
2 & 62 & 25.0\,\% (5/20) & 0.0028 & 0.004 & +6.47\,\%  & +0.86\,\% \\
3 & 54 & 31.6\,\% (6/19) & 0.0037 & 0.008 & +8.23\,\%  & +1.19\,\% \\
4 & 42 & 44.4\,\% (8/18) & 0.0040 & 0.010 & +20.83\,\% & +0.51\,\% \\
5 & 78 & 13.6\,\% (3/22) & 0.0020 & 0.002 & +2.93\,\%  & +1.28\,\% \\
6 & 42 & 37.5\,\% (6/16) & 0.0043 & 0.010 & +16.07\,\% & +0.25\,\% \\
7 & 34 & 42.9\,\% (6/14) & 0.0037 & 0.006 & +13.01\,\% & +0.23\,\% \\
8 & 78 & 9.5\,\% (2/21)  & 0.0020 & 0.002 & +4.62\,\%  & +1.28\,\% \\
\bottomrule
\end{tabular}%
}
\caption{Secret-key rate (SKR) enhancement from optimal tuned $p_{LR}$, with respect to the one computed in absence of local randomization, for each asymmetric configuration. The second column indicates the minimum distance for which $p_{LR} >0$, and the third, the number of simulated points (as percentage and as ratio of number of points divided by total number of points) where the local randomization provides an improvement.}
\label{tab:qbf_comparison}
\end{table}

\begin{figure}[!h]
    \centering
    \includegraphics[width=1.0\linewidth]{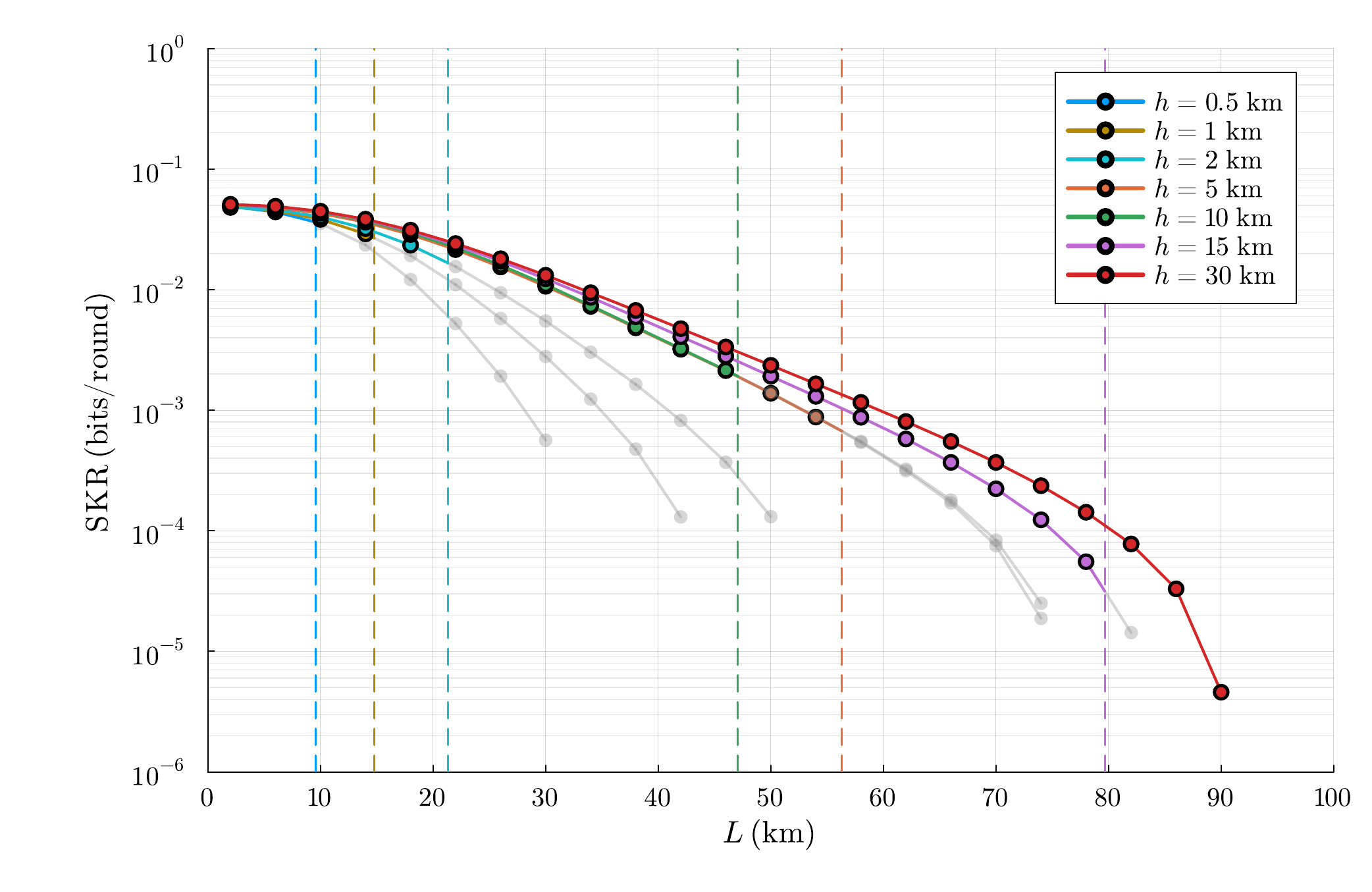}
    \caption{Secret-key rate (SKR) as a function of the total Alice--Bob distance $L=L_A+L_B$ for the asymmetric MDI QKD configuration $5$, under ratios $L_A/L_B=4/3$ and $\gamma_A/\gamma_B=5/4$, at different altitudes $h$. The block size is set to $n=10^8$. Vertical dashed lines indicate the maximum communication distances compatible with the weak-turbulence conditions. Points marked in gray are beyond these limits and therefore excluded from the physically valid regime. The coherent-state amplitudes are optimized subject to the fixed ratio, at each operating point. All other parameters are reported in Table~\ref{tab:simulation_parameters}.}
    \label{fig:asym_turb}
\end{figure}

Having identified the configuration $5$ as one of the most favorable asymmetric arrangements, we next examine its dependence on the propagation altitude, for $n=10^8$. Fig.~\ref{fig:asym_turb} shows that, as in the symmetric setup, lower altitudes generally lead to stronger attenuation and turbulence, reducing both the achievable SKR and the maximum communication distance. At higher altitudes, atmospheric effects become progressively weaker, and the performance approaches the atmosphere-free regime. 
As already observed in Fig.~\ref{fig:SecondFigure}, the curves at $h=5\,\mathrm{km}$ and $h=10\,\mathrm{km}$ exhibit a crossing, consistent with the non-monotonic altitude dependence of the Hufnagel--Valley turbulence profile discussed in Subsec. \ref{subsec:symm} . The SKRs reported here are generally higher than those in the corresponding symmetric configuration shown in Fig.~\ref{fig:SecondFigure}, mainly because the present simulations employ a larger block size ($n=10^8$ instead of $n=10^6$), which reduces finite-size penalties. As in the symmetric analysis, the vertical boundaries indicate the applicability limits of the adopted weak-turbulence model, beyond which the corresponding results should not be regarded as physically reliable predictions.

\begin{figure}[!h]
    \centering
    \includegraphics[width=1\linewidth]{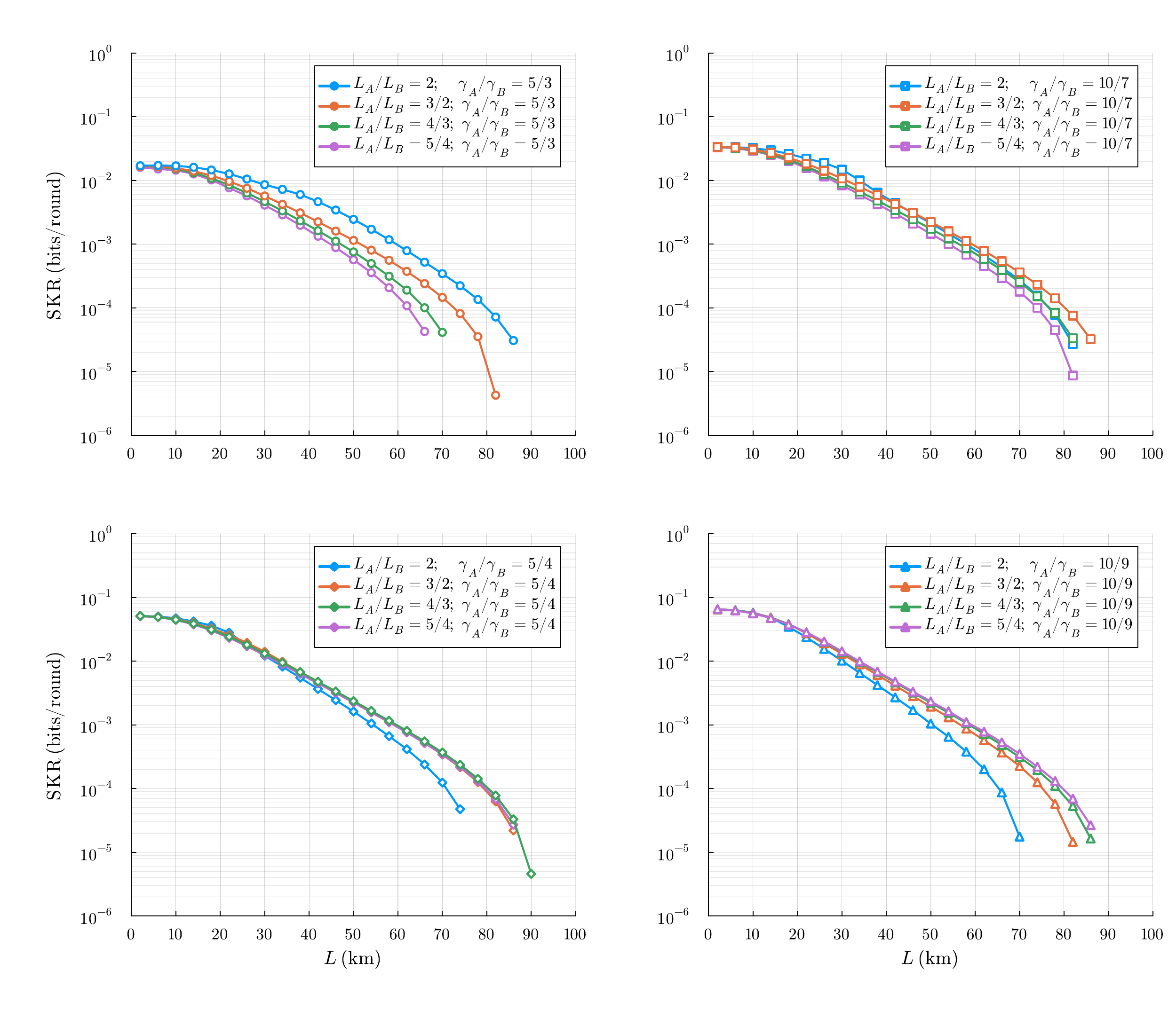}
    \caption{
    Secret-key rate (SKR) as a function of the total Alice--Bob distance $L=L_A+L_B$ for asymmetric MDI QKD configurations with $L_A>L_B$ and $\gamma_A>\gamma_B$, considering different ratios of arm lengths $L_A/L_B$ and transmitted coherent-state amplitudes $\gamma_A/\gamma_B$. The simulations are performed at $h=30\,\mathrm{km}$ in the atmosphere-free regime and a fixed block size $n=10^8$. The coherent-state amplitudes are optimized subject to the specified ratios, at each operating point. The remaining simulation parameters are reported in Table~\ref{tab:simulation_parameters}. }
    \label{fig:diff_ratios}
\end{figure}

Finally, we investigate how the degree of asymmetry affects the SKR by varying the ratios $L_A/L_B$ and $\gamma_A/\gamma_B$. Fig. \ref{fig:diff_ratios} shows that the relative ordering of the SKR curves depends not only on the selected ratios but also on the total communication distance. Here we fixed $n=10^8$ and $h=30\,\text{km}$. This behavior can again be understood from $\langle\mu_{\mathrm{dark}}\rangle$, since the balance between the fields arriving at the relay is determined by the products $\sqrt{\eta_A}\gamma_A$ and $\sqrt{\eta_B}\gamma_B$, rather than by the transmitted amplitudes alone. In particular, for a fixed arm-length ratio $L_A/L_B>1$, minimizing $\langle\mu_{\mathrm{dark}}\rangle$ with respect to $\gamma_A$, while keeping $\gamma_B$ fixed, yields
\begin{equation}
\left(\frac{\gamma_A}{\gamma_B}\right)_{\mathrm{bal}}
=
\frac{\langle\sqrt{\eta_A\eta_B}\rangle}{\langle\eta_A\rangle}
\simeq
\sqrt{\frac{\langle\eta_B\rangle}{\langle\eta_A\rangle}}.
\end{equation}
The approximation holds when the relative fluctuations of the channel transmittances are sufficiently small.
\footnote{In the atmosphere-free regime considered here, transmittance fluctuations arise exclusively from pointing errors. For our parameters, $\theta_{\mathrm{pe}}=1\,\mu\mathrm{rad}$ is substantially smaller than the beam divergence $\lambda/(\pi w_0)\simeq5.1\,\mu\mathrm{rad}$. A Gaussian approximation for the collection efficiency in the far field indicates that replacing $\langle\sqrt{\eta_i}\rangle$ with $\sqrt{\langle\eta_i\rangle}$ introduces a correction of approximately $0.3\%$ per channel, supporting the use of this approximation to interpret the numerical trends.} Under the atmosphere-free conditions considered in Fig.~\ref{fig:diff_ratios}, diffraction determines the beam width, while pointing errors introduce fluctuations in the collection efficiency. In the far-field regime, Eqs.~\eqref{eq:horizontal_diffraction} and~\eqref{eq:far_collection}, together with Eq.~\eqref{eq:atmosphere_free_beam}, imply that the diffraction-limited collection efficiency scales approximately as $L^{-2}$. Since the pointing parameters are identical for both channels, their mean transmittances approximately follow the same scaling, leading to $(\gamma_A/\gamma_B)_{\mathrm{bal}}\simeq L_A/L_B$.
\footnote{For the parameters adopted in our simulations, the Rayleigh range is $L_R=\pi w_0^2/\lambda\simeq9.8\,\mathrm{km}$. Hence, the far-field approximation is most appropriate for the longer communication distances considered in Fig.~\ref{fig:diff_ratios}, where the individual arm lengths exceed $L_R$ by several times. It is less accurate at shorter distances.} 

This relation provides a quantitative interpretation of the trends observed across the four panels of Fig.~\ref{fig:diff_ratios}. For larger arm-length ratios, such as $L_A/L_B=2$, greater transmitted amplitude asymmetry is required to compensate for the unequal channel attenuation, and reducing $\gamma_A/\gamma_B$ generally leads to lower SKRs, particularly at longer communication distances. Conversely, for smaller arm-length ratios, such as $L_A/L_B=5/4$, excessively large amplitude ratios produce an imbalance between the received fields, reducing the achievable SKR. Although the balancing condition minimizes the residual dark-port intensity rather than directly maximizing the SKR, it explains the main dependence of the observed performance on the combined distance and amplitude asymmetries.

\newpage

\section{Discussion}\label{Sec:Discussion}
In this work, we have presented the first application of the finite-size, general security framework from the MEAT on an MDI QKD protocol combining imperfect source characterization and free-space quantum communication. Our variable-length framework provides security against general attacks, incorporating source imperfections through fidelity constraints and evaluating the resulting Rényi entropic quantities by means of conic optimization \cite{arqand2025MEAT,pereira2025optimalkeyratesquantum,Navarro2025, Navarro_finite}. 
Our framework also extends previous numerical MEAT implementations \cite{kamin25MEATsecurity,tupkary2026rigorouscompletesecurityproof,wiesemann2026} by allowing probabilistic key generation conditioned on the relay outcome. Besides removing the requirement of a fixed overall testing probability, this reduces QRNG consumption, since private randomness for the key/testing selection is only required for eligible rounds.

To characterize the horizontal free-space links, we have combined atmospheric extinction, evaluated using LOWTRAN \cite{Kneizys1988LOWTRAN,PicciaLowtran}, with the Hufnagel--Valley turbulence model \cite{Hufnagel1974,Valley1980,AndrewsLaserBeam}. Diffraction, turbulence-induced beam broadening and wandering, and pointing fluctuations are incorporated into the probability distribution of transmittance following the approaches of~\cite{recolons2007beamwander,vasylyev2012toward,pirandola2021limits}. This description covers atmospheric propagation within the weak-turbulence regime and the atmosphere-free limit, where pointing fluctuations remain relevant.

Our numerical results highlight the impact of finite-size effects and atmospheric conditions on the achievable secret-key rate. In symmetric configurations, increasing the block size improves the SKR and extends the communication range, with positive rates obtained even for $n=3\times10^4$ rounds under the considered conditions. For $n=10^8$, positive SKRs are obtained at a total distance of $82$ km in the atmosphere-free regime, particularly for asymmetric setups in which the larger transmitted amplitude is assigned to the longer arm, namely for configurations 5 and 8.

The altitude-dependent analysis further reveals the competing effects of atmospheric extinction and turbulence. For asymmetric configurations, we have shown that the interplay between communication distances and transmitted coherent-state amplitudes is crucial. In particular, assigning a larger amplitude to the more attenuated channel can partially compensate for the imbalance between the optical fields arriving at the relay, improving the achievable SKR.

Previous MDI QKD studies have achieved remarkable results in terms of communication distance and finite-size performance. For instance, decoy-state MDI QKD has been experimentally demonstrated over $442\,\mathrm{km}$ of ultralow-loss optical fiber \cite{Liu2023FiveIntensity}, while a free-space implementation has achieved $19.2\,\mathrm{km}$ over an urban atmospheric channel \cite{Cao2020FreeSpaceMDI}. Regarding finite-size security, \cite{Curty2014FiniteKey} demonstrated the feasibility of fiber-based MDI QKD with blocks of $10^{12}$--$10^{14}$ transmitted signals under general attacks, while CV MDI QKD analyses considering Gaussian attacks have reported positive key rates for blocks between $10^6$ and $10^9$ data points \cite{Papanastasiou2017FiniteSize}. In contrast with these works, we achieve SKRs for smaller blocks under favorable atmosphere-free conditions. However, the differences in encoding schemes, channel models, source assumptions, and security frameworks prevent a meaningful comparison in terms of either distance or SKR yield.

We have also investigated the effect of local randomization \cite{Renner2005}, observing a markedly different behavior from that reported for BB84 and six-state protocols in ~\cite{staffieri2026}. For those instances, optimized preprocessing noise substantially improves the SKR at high qubit error rates and extends the maximum tolerable disturbance, also in the finite-size regime. In contrast, its contribution in the present MDI QKD implementation is considerably more limited. Local randomization is generally not advantageous in symmetric configurations, whereas asymmetric setups exhibit improvements at selected operating points, with small optimal preprocessing probabilities, typically of the order of $10^{-3}$--$10^{-2}$. Although appreciable relative enhancements can occur, the overall benefit is much less pronounced than in the BB84 and six-state protocols.

Several directions remain open for future investigations. A more detailed treatment of imperfect state preparation could explicitly incorporate phase and amplitude mismatches between the coherent states prepared by Alice and Bob, complementing the fidelity-based source characterization adopted here \cite{wiesemann2026, pereira2025optimalkeyratesquantum}. This characterization could also be expanded to explicitly account for side channels such as Trojan-horse attacks \cite[Appendix D]{wiesemann2026}, or source-detector correlations \cite{tupkary2026meatsourcedetector}.

Extending the horizontal-channel model beyond weak turbulence would allow the investigation of additional propagation regimes. Furthermore, including aerosol extinction and a more detailed description of background radiation would enable the analysis of specific atmospheric conditions, accounting for geographical location, seasonal variations, and daytime or nighttime operation. Another natural extension is the study of slant links between platforms operating at different altitudes, where atmospheric properties vary along the propagation path. This would result in a more realistic model, particularly for metropolitan networks where the nodes are distributed at different heights.

Finally, another relevant development is the extension of the present framework to other discrete-variable (DV) QKD protocols, particularly those employing decoy states \cite{tupkary2025qkdsecurityproofsdecoystate,kamin25MEATsecurity}. Incorporating multiple intensity settings would allow a more detailed characterization of the photon-number contributions to the detection statistics.  Combined with the free-space channel model developed here, this extension would enable the investigation of decoy-state MDI QKD protocols under atmospheric propagation and asymmetric channel conditions.

\section*{Acknowledgments}

This project has received funding from the European Union’s Digital Europe Programme under the project QUARTERNEXT (101305103), and the European Innovation Council's (EIC) Horizon Europe Accelerator Programme under the project MIQRO (101161539).  G.S. acknowledges financial support from the Italian Space Agency (agreement n. 2024-60-HH.0) and thanks Cosmo Lupo for his valuable guidance, insightful discussions, and continuous support throughout this research. We thank the Luxquanta S.L. team, J. Cabanillas and A. Cabanillas for their daily support and encouragement, Mateus Araújo for insightful indications on numerical methods for QKD; Jerome Wiesemann and Devashish Tupkary for fruitful discussions and suggestions on the MEAT framework; and Michail Karatarakis and Axiomatic AI for their inputs on formalizing the security proof.

\section*{Code availability}
The numerical simulations described in Sections \ref{Sec:numerics} and \ref{Sec:Results} make use of the software LOWTRAN7 \cite{Kneizys1988LOWTRAN} and its Python implementation from the repository \cite{PicciaLowtran}. We further used the programming language Julia \cite{JuliaLang} for all simulations, and in particular the optimization suite JuMP \cite{Lubin2023} together with the conic solver Hypatia \cite{coey2022performance,coey2022solving} and its extension ConicQKD \cite{Navarro2025}. The particular codes employed for elaborating the plots are available upon reasonable request.

\printbibliography

\appendix

\section{Derivation of the honest probability distribution}
\label{app:q_honest}
In this Appendix we derive the honest conditional
probabilities $\widetilde q(c|x,y)$ used to construct
the reference probability distribution $q_{\hat C}$
introduced in Sec.~\ref{Sec:numerics}. We assume statistically independent fluctuations of the Alice's and Bob's free-space links, described by the probabilities $F_{\mathrm H_A}(\eta_A)$ and $F_{\mathrm H_B}(\eta_B)$, respectively. 

Let $\gamma_x$ and $\beta_y$ denote the coherent amplitudes associated with the reference preparations of Alice and Bob, respectively. In particular the vacuum preparation is included by setting $\gamma_x=0$ or $\beta_y=0$. For fixed channel transmittances $\eta_A$ and $\eta_B$ the joint state arriving at the repeater is 
\begin{align}
\rho_{xy}(\eta_A,\eta_B)
={}&
\ket{\sqrt{\eta_A}\gamma_x}
\bra{\sqrt{\eta_A}\gamma_x}_{A}
\otimes
\ket{\sqrt{\eta_B}\beta_y}
\bra{\sqrt{\eta_B}\beta_y}_{B}.
\label{eq:fixed_channel_state_app}
\end{align}
The two incoming modes interfere at a balanced beam splitter for which we adopt the convention
\begin{align}
U_{\rm BS}
\ket{\gamma}_{A}
\ket{\beta}_{B}
=
\ket{\frac{\gamma+\beta}{\sqrt{2}}}_{ D0}
\ket{\frac{\gamma-\beta}{\sqrt{2}}}_{D1 },
\label{eq:BS_convention_app}
\end{align}
where the subscripts $D0$ and $D1$ identify the two output modes. Therefore,
\begin{align}
U_{\rm BS}
\rho_{xy}(\eta_A,\eta_B)
U_{\rm BS}^{\dagger}
=
\pro{\zeta}_{D 0}
\otimes
\pro{\xi}_{D 1}
\label{eq:BS_output_state_app}
\end{align}
where
\begin{align}
\zeta
&=
\frac{
\sqrt{\eta_A}\gamma_x
+
\sqrt{\eta_B}\beta_y
}{\sqrt{2}},
\label{eq:zeta_app}
\\
\xi
&=
\frac{
\sqrt{\eta_A}\gamma_x
-
\sqrt{\eta_B}\beta_y
}{\sqrt{2}}.
\label{eq:xi_app}
\end{align}
We define the corresponding mean photon numbers at the two output ports as
\begin{align}
\mu_ {0}
=
|\zeta|^2,
\qquad
\mu_{1}
=
|\xi|^2.
\label{eq:mu_definition_app}
\end{align}
Explicitly,
\begin{align}
\mu_{0}
={}&
\frac{1}{2}
\Big[
\eta_A|\gamma_x|^2
+
\eta_B|\beta_y|^2
\nonumber\\
&\hspace{1.5cm}
+
2\sqrt{\eta_A\eta_B}\,
\operatorname{Re}
\left(
\gamma_x
\beta_y^{*}
\right)
\Big],
\label{eq:mu0_app}
\\
\mu_{1}
={}&
\frac{1}{2}
\Big[
\eta_A|\gamma_x|^2
+
\eta_B|\beta_y|^2
\nonumber\\
&\hspace{1.5cm}
-
2\sqrt{\eta_A\eta_B}\,
\operatorname{Re}
\left(
\gamma_x
\beta_y^{*}
\right)
\Big].
\label{eq:mu1_app}
\end{align}
Before including dark counts, the two detectors define the four-outcome POVM
\begin{align}
    P^{\rm nc}
    &=
    \pro{0}_{D0}
    \otimes
    \pro{0}_{D1}
    \label{eq:Pnc_app}
    \\
    P^0
    &=
    \left(
    \mathbb{I}_{D0}-\pro{0}_{D0}
    \right)
    \otimes
    \pro{0}_{D1},
    \label{eq:P0_app}
    \\
    P^1
    &=
    \pro{0}_{D0}
    \otimes
    \left(
    \mathbb{I}_{D1}-\pro{0}_{D1}
    \right),
    \label{eq:P1_app}
    \\
    P^{\rm dc}
    &=
    \left(
    \mathbb{I}_{D0}-\pro{0}_{D0}
    \right)
    \otimes
    \left(
    \mathbb{I}_{D1}-\pro{0}_{D1}
    \right),
    \label{eq:Pdc_app}
\end{align}
corresponding, respectively, to a no-click event, a click only at detector $0$,  a click only at detector $1$ and a double click. For fixed values of the transmittances we define
\begin{align}
    \pi_\nu(x,y, \eta_A, \eta_B) =
    \Tr \left\{
    P^\nu\,
    U_{BS}\,
    \left[
    \rho_{xy}(\eta_A,\eta_B)
    \right]
    U_{BS}^\dagger
    \right\}
    \label{eq:pi_definition_app}
\end{align}
where  $\nu\in\{{\rm nc},0,1,{\rm dc}\}$. Using 
\begin{align}
\left|\langle0|\zeta\rangle\right|^2
&=
e^{-|\zeta|^2}
=
e^{-\mu_{0}},
&
\left|\langle0|\xi\rangle\right|^2
&=
e^{-|\xi|^2}
=
e^{-\mu_{1}},
\end{align}
the four probabilities are
\begin{align}
\pi_{\rm nc}
&=
e^{-\mu_0-\mu_1},
\label{eq:pi_nc_app}
\\
\pi_0
&=
\left(1-e^{-\mu_0}\right)
e^{-\mu_1},
\label{eq:pi0_app}
\\
\pi_1
&=
e^{-\mu_0}
\left(1-e^{-\mu_1}\right),
\label{eq:pi1_app}
\\
\pi_{\rm dc}
&=
\left(1-e^{-\mu_0}\right)
\left(1-e^{-\mu_1}\right).
\label{eq:pi_dc_app}
\end{align}

We now include the trusted dark-count probabilities $p_d^0$ and $p_d^1$ associated with detectors $D0$ and $D1$, respectively. According to the protocol definition, both no-click and double-click events are assigned to inconclusive outcome $\perp$. The effective three-outcome POVM is obtained from
\begin{align}
\begin{pmatrix}
M^\perp\\
M^0\\
M^1
\end{pmatrix}
=
\begin{pmatrix}
1-(1-p_d^0)p_d^1-(1-p_d^1)p_d^0
&
p_d^1
&
p_d^0
&
1
\\
p_d^0(1-p_d^1)
&
1-p_d^1
&
0
&
0
\\
(1-p_d^0)p_d^1
&
0
&
1-p_d^0
&
0
\end{pmatrix}
\begin{pmatrix}
P^{\rm nc}\\
P^0\\
P^1\\
P^{\rm dc}
\end{pmatrix},
\label{eq:D_relation_app}
\end{align}
For fixed channel transmittances, the probability of the effective outcome $c\in \{\perp,0,1\}$ is therefore
\begin{align}
     p(c|x,y, \eta_A, \eta_B) =
    \Tr \left[
    M^c\,
    U_{BS}\,
    \left[
    \rho_{xy}(\eta_A,\eta_B)
    \right]
    U_{BS}^\dagger
    \right]
    \label{eq:fixed_probability_app}
\end{align}
Using Eqs. \eqref{eq:pi_nc_app}--\eqref{eq:pi_dc_app} and the transformation in Eq. \eqref{eq:D_relation_app}, the probability of the outcome $0$ is 
\begin{align}
p(0|x,y,\eta_A,\eta_B)
=&\,
p_d^0(1-p_d^1)\pi_{\rm nc}
+
(1-p_d^1)\pi_0
\nonumber\\
=&\,
(1-p_d^1)e^{-\mu_1}-
(1-p_d^0)(1-p_d^1)
e^{-\mu_0-\mu_1.}
\label{eq:p0_fixed_app}
\end{align}
Analogously
\begin{align}
p(1|x,y,\eta_A,\eta_B)
=&\,
(1-p_d^0)p_d^1\pi_{\rm nc}
+
(1-p_d^0)\pi_1
\nonumber\\
=&\,
(1-p_d^0)e^{-\mu_0}
-
(1-p_d^0)(1-p_d^1)
e^{-\mu_0-\mu_1}.
\label{eq:p1_fixed_app}
\end{align}
The inconclusive probability may be obtained either directly from the
first row of linear transformation in \eqref{eq:D_relation_app} or from normalization. This gives
\begin{align}
p(\perp|x,y,\eta_A,\eta_B)
={}&
1
-
(1-p_d^0)e^{-\mu_0}
-
(1-p_d^1)e^{-\mu_1}
\nonumber\\
&+
2(1-p_d^0)(1-p_d^1)
e^{-\mu_0-\mu_1}.
\label{eq:pperp_fixed_app}
\end{align}
We finally average  the fixed-transmittance probabilities over the fluctuations of the two free-space links. 
\begin{align}
  \widetilde q(c|x,y)
    =
    \int_0^{\eta_A^{\max}} d\eta_A
    \int_0^{\eta_B^{\max}} d\eta_B\,
    F_{\mathrm H_A}(\eta_A)
    F_{\mathrm H_B}(\eta_B)\,
    p(c|x,y,\eta_A,\eta_B), \qquad c \in \{0,1,\perp\}
\end{align}
To simplify the resulting expressions, we introduce the PDT-averaged quantities
\begin{align}
\mathcal J_0(x,y)
:={}&
\int_0^{\eta_A^{\max}}
d\eta_A
\int_0^{\eta_B^{\max}}
d\eta_B\,
F_{\rm H_A}(\eta_A)
F_{\rm H_B}(\eta_B)
e^{-\mu_0},
\label{eq:J0_MDI}
\\
\mathcal J_1(x,y)
:={}&
\int_0^{\eta_A^{\max}}
d\eta_A
\int_0^{\eta_B^{\max}}
d\eta_B\,
F_{\rm H_A}(\eta_A)
F_{\rm H_B}(\eta_B)
e^{-\mu_1},
\label{eq:J1_MDI}
\\
\mathcal J_{\rm nc}(x,y)
:={}&
\int_0^{\eta_A^{\max}}
d\eta_A
\int_0^{\eta_B^{\max}}
d\eta_B\,
F_{\rm H_A}(\eta_A)
F_{\rm H_B}(\eta_B)
e^{-\mu_0-\mu_1}.
\label{eq:Jnc_MDI}
\end{align}
Since in the sum $\mu_0+\mu_1$ the interference terms cancel out, we note that Eq. \eqref{eq:Jnc_MDI} can be factorized into 
\begin{align}
\mathcal J_{\rm nc}(x,y)
={}&
\left[
\int_0^{\eta_A^{\max}}
d\eta_A\,
F_{\rm H_A}(\eta_A)
e^{-\eta_A|\gamma_x|^2}
\right]
\times
\left[
\int_0^{\eta_B^{\max}}
d\eta_B\,
F_{\rm H_B}(\eta_B)
e^{-\eta_B|\beta_y|^2}
\right].
\label{eq:Jnc_factorized_MDI}
\end{align}
In contrast, $\mathcal J_0(x,y)$ and $\mathcal J_1(x,y)$ do not generally factorize, since the interference term in Eqs.~\eqref{eq:mu0_app} and
\eqref{eq:mu1_app} couples the two channel transmittances through $\sqrt{\eta_A\eta_B}$. The honest conditional probabilities are consequently
\begin{align}
\widetilde q(0|x,y)
=&\,
(1-p_d^1)\,
\mathcal J_1(x,y)
-
(1-p_d^0)(1-p_d^1)\,
\mathcal J_{\rm nc}(x,y),
\label{eq:q0_integrated_app}
\\
\widetilde q(1|x,y)
=&\,
(1-p_d^0)\,
\mathcal J_0(x,y)
-
(1-p_d^0)(1-p_d^1)\,
\mathcal J_{\rm nc}(x,y),
\label{eq:q1_integrated_app}
\\
\widetilde q(\perp|x,y)
=&\,
1
-
(1-p_d^0)\,
\mathcal J_0(x,y)
-
(1-p_d^1)\,
\mathcal J_1(x,y)
\nonumber\\
&+
2(1-p_d^0)(1-p_d^1)\,
\mathcal J_{\rm nc}(x,y).
\label{eq:qperp_integrated_app}
\end{align}

\section{Local randomization at Alice's side} \label{App:NoisyPreproc}
This appendix provides the complete construction of the key and pinching maps presented in Sec.~\ref{Sec:numerics} in the presence of local randomization on Alice's raw-key variable, together with the facial reduction used in the numerical implementation.
The choice of Alice rather than Bob follows from the
adopted direct-reconciliation scheme, in which Alice's
raw-key string defines the reference key to be reconstructed
by Bob. Accordingly, local randomization is applied
to Alice's raw key register
\cite{Renner2005,Kraus2005}. More concretely, after a key-generation round Alice replaces her raw bit $X \in \{0,1\}$ by 
\begin{align}
    S = X\oplus F, \qquad F \sim \mathrm{Bern}(p_{\mathrm{LR}})
\end{align}
where $p_{\mathrm{LR}}\in (0,1/2]$ is the trusted bit-flip probability and $F$ is generated locally and independently from Eve. 

\subsection{Noiseless maps.}

We first introduce the key and pinching maps in the absence of noisy
preprocessing. Let
\begin{align}
\Pi^{\mathrm K}_{B}
:=
\sum_{y=0,1}
\pro{y}_{B},
\end{align}
and define the two orthogonal key-generation sectors
\begin{align}
Z_k
:=
\pro{k}_{A}
\otimes
\Pi^{\mathrm K}_{B},
\qquad
k=0,1.
\label{eq:Zj-noisy-app}
\end{align}
They satisfy $Z_0 Z_1 =0,$ and $Z_0+Z_1
=T$, where
\begin{align}
T
:=
\sum_{x,y=0,1}
\pro{x}_{A}
\otimes
\pro{y}_{B}
\label{eq:G-original-app}
\end{align}
is the projector onto the key-generation subspace. To retain a coherent representation of Alice's raw-key
measurement before dephasing, we introduce a two-dimensional
auxiliary register $R$, with computational basis
$\{\ket{0}_{R},\ket{1}_{R}\}$.
This register coherently encodes the key value, whose
classical counterpart is the register $S$ introduced in
the protocol description, obtained by measuring $R$
in the computational basis. In the absence of noisy
preprocessing, this value coincides with Alice's raw bit. The noiseless key-map operator is
\begin{align}
K
:=
\sum_{k=0,1}
\ket{k}_{R}
\otimes
Z_k.
\label{eq:K-noiseless-MDI}
\end{align}
The corresponding key map is
\begin{align}
\mathcal G(\cdot)
:=
K (\cdot) K^\dagger.
\label{eq:G-noiseless-MDI}
\end{align}
As the two relay announcements $c'=0,1$ are treated symmetrically, $\mathcal G^0
=
\mathcal G^1
=
\mathcal G$.
The corresponding pinching map acts on the auxiliary register $R$.
Defining
\begin{align}
Q_k
:=
\pro{k}_{R}
\otimes
T,
\qquad
k=0,1,
\label{eq:Q-noiseless}
\end{align}
we write
\begin{align}
\mathcal Z(\cdot)
:=
\sum_{k=0,1}
Q_k (\cdot) Q_k.
\label{eq:Z-noiseless}
\end{align}

\subsection{Local randomization.} 

The trusted randomization can be absorbed into the effective
measurement defining the raw key. In analogy with a binary symmetric
classical channel, we introduce the two noisy POVM elements
\begin{align}
    \Lambda_{0}^{(p_{\mathrm{LR}})}
    &=
    (1-p_{\mathrm{LR}}) Z_0 + p_{\mathrm{LR}} Z_1,
    \label{eq:Lambda0-MDI}
    \\
    \Lambda_{1}^{(p_{\mathrm{LR}})}
    &=
    p_{\mathrm{LR}} Z_0 + (1-p_{\mathrm{LR}}) Z_1.
    \label{eq:Lambda1-MDI}
\end{align}
Since $Z_0$ and $Z_1$ are mutually orthogonal projectors, the
positive square roots of the noisy POVM elements can be written
explicitly as
\begin{align}
    \sqrt{\Lambda_{0}^{(p_{\mathrm{LR}})}}
    &=
    \sqrt{1-p_{\mathrm{LR}}}\,Z_0
    +
    \sqrt{p_{\mathrm{LR}}}\,Z_1,
    \label{eq:sqrt-Lambda0-MDI}
    \\
    \sqrt{\Lambda_{1}^{(p_{\mathrm{LR}})}}
    &=
    \sqrt{p_{\mathrm{LR}}}\,Z_0
    +
    \sqrt{1-p_{\mathrm{LR}}}\,Z_1.
    \label{eq:sqrt-Lambda1-MDI}
\end{align}
In the presence of local randomization, we use
the same auxiliary register $R$ to coherently encode
the preprocessed key value $S=X\oplus F$, rather than
Alice's original raw bit $X$. The 
key-map operator reads
\begin{align}
    K^{(p_{\mathrm{LR}})}
    :=
    \sum_{r=0,1}
    \ket{r}_{R}
    \otimes
    \sqrt{\Lambda_{r}^{(p_{\mathrm{LR}})}} .
    \label{eq:K-noisy-MDI}
\end{align}
The corresponding key map is
\begin{align}
    \mathcal G^{(p_{\mathrm{LR}})}(\cdot)
    :=
    K^{(p_{\mathrm{LR}})} (\cdot) K^{(p_{\mathrm{LR}})\dagger}.
    \label{eq:G-noisy-MDI}
\end{align}
As in the noiseless implementation, the two relay announcements
$c'=0,1$ are treated symmetrically, and therefore
$ \mathcal G^{0,(p_{\mathrm{LR}})}
    =
    \mathcal G^{1,(p_{\mathrm{LR}})}
    =
    \mathcal G^{(p_{\mathrm{LR}})}$.
For later convenience, the action of the map can be written
explicitly as
\begin{align}
    \mathcal G^{(p_{\mathrm{LR}})}(\cdot)
    =
    \sum_{r,t=0,1}
    \ket{r}\!\bra{t}_{R}
    \otimes
    \sqrt{\Lambda_{r}^{(p_{\mathrm{LR}})}}
    (\cdot)
    \sqrt{\Lambda_{t}^{(p_{\mathrm{LR}})}}.
    \label{eq:G-expanded-noisy}
\end{align}
In particular, the off-diagonal terms $r\neq t$ retain the coherence
between the two possible raw-key values before the final classical
key register is produced.
Finally the pinching map reads the same as in the noiseless implementation, its action on the output of the key map is therefore
\begin{align}
    \mathcal Z
    \circ
    \mathcal G^{(p_{\mathrm{LR}})}(\cdot)
    =
    \sum_{r=0,1}
    \pro{r}_{R}
    \otimes
    \sqrt{\Lambda_{r}^{(p_{\mathrm{LR}})}}
    (\cdot)
    \sqrt{\Lambda_{r}^{(p_{\mathrm{LR}})}}.
    \label{eq:ZG-noisy}
\end{align}
Thus, the pinching removes only the coherences between the two values of the preprocessed raw-key register, while leaving unchanged the corresponding conditional blocks produced by the key map, which may still retain coherences between Alice's original key sectors.

\subsection{Facial reduction}

The maps in  Eqs.~\eqref{eq:G-noisy-MDI} and
\eqref{eq:ZG-noisy} have outputs in the nominal
$18$-dimensional Hilbert space
\begin{align}
    \mathcal H_{R}\otimes
    \mathcal H_{A}\otimes
    \mathcal H_{B},
\end{align}
since $\dim\mathcal H_{R}=2$ and
$\dim\mathcal H_{A}=\dim\mathcal H_{B}=3$. However, their outputs are necessarily supported on lower-dimensional subspaces. It is therefore convenient to perform a facial reduction before introducing them into the conic program.

Let $\Phi:
\mathcal L(\mathcal H_{\mathrm{in}})
\longrightarrow
\mathcal L(\mathcal H_{\mathrm{out}})$ be a completely positive map, and let $r_\Phi
:=\mathrm{rank}\left[\Phi(\mathbb I_{\mathrm{in}})\right]$. We call
\begin{align}
W_\Phi:
\mathbb C^{r_\Phi}
\longrightarrow
\mathcal H_{\mathrm{out}}
\end{align}
a facial-reduction isometry for $\Phi$ if its columns form an
orthonormal basis of
$\mathrm{supp}[\Phi(\mathbb I_{\mathrm{in}})]$. 
Equivalently, $W_\Phi$
satisfies
\begin{align}
W_\Phi^\dagger W_\Phi
&=
\mathbb{I}_{r_\Phi},
\label{eq:FR-isometry-condition}
\\
W_\Phi W_\Phi^\dagger
&=
\Pi_{\mathrm{supp}[\Phi(\mathbb I_{\mathrm{in}})]},
\label{eq:FR-support-condition}
\end{align}
where
$\Pi_{\mathrm{supp}[\Phi(\mathbb I_{\mathrm{in}})]}$
denotes the orthogonal projector onto the support of
$\Phi(\mathbb I_{\mathrm{in}})$.
The relevance of $\Phi(\mathbb I_{\mathrm{in}})$ follows directly from the
Kraus representation. If

\begin{align}
\Phi(\cdot)
=
\sum_s K_s (\cdot) K_s^\dagger,
\end{align}
then
\begin{align}
\Phi(\mathbb I_{\mathrm{in}})
=
\sum_s K_s K_s^\dagger,
\end{align}
and hence
\begin{align}
\mathrm{supp}[\Phi(\mathbb I_{\mathrm{in}})]
=
\mathrm{span}_s[
\mathrm{Ran}(K_s)].
\label{eq:FR-support-Kraus}
\end{align}
Therefore, the support of $\Phi(\mathbb I_{\mathrm{in}})$ is precisely the
subspace of $\mathcal H_{\mathrm{out}}$ that can be populated by the
map, while its rank gives the dimension of the corresponding effective
output space. Once $W_\Phi$ is chosen, the map can be represented on
this reduced output space through the compression
\begin{align}
\widehat{\Phi}(\cdot)
:=
W_\Phi^\dagger[\Phi(\cdot)]W_\Phi,
\end{align}
with
\begin{align}
\Phi(\cdot)
=
W_\Phi
[\widehat{\Phi}(\cdot)]
W_\Phi^\dagger.
\end{align}

We now apply this construction separately to
$\mathcal G^{(p_{\mathrm{LR}})}$ and
$\mathcal Z\circ\mathcal G^{(p_{\mathrm{LR}})}$, since the pinching changes
the support of the output.

\subsubsection{Facial reduction of $\mathcal G^{(p_{\mathrm{LR}})}$}

The noisy key map has the single-Kraus form
\begin{align}
\mathcal G^{(p_{\mathrm{LR}})}(\cdot)
=
K^{(p_{\mathrm{LR}})}(\cdot)K^{(p_{\mathrm{LR}})\dagger},
\end{align}
where
\begin{align}
K^{(p_{\mathrm{LR}})}:
\mathcal H_{A}\otimes\mathcal H_{B}
\longrightarrow
\mathcal H_{R}\otimes
\mathcal H_{A}\otimes\mathcal H_{B}.
\end{align}
Therefore,
\begin{align}
\mathcal G^{(p_{\mathrm{LR}})}(\mathbb I_{AB})
=
K^{(p_{\mathrm{LR}})}K^{(p_{\mathrm{LR}})\dagger},
\end{align}
and, since
$\mathrm{supp}(KK^\dagger)=\mathrm{Ran}(K)$,
\begin{align}
\mathrm{supp}[
\mathcal G^{(p_{\mathrm{LR}})}(\mathbb I_{AB})]
=
\mathrm{Ran}(K^{(p_{\mathrm{LR}})}).
\label{eq:supp-G-ranK}
\end{align}

To determine this range, it is sufficient to evaluate
$K^{(p_{\mathrm{LR}})}$ on a basis of its input space, since by linearity the
range of an operator is the span of the images of any basis of its
domain. Define
\begin{align}
\ket{v_0}^{(p_{\mathrm{LR}})}_{R}
&=
\sqrt{1-p_{\mathrm{LR}}}\ket{0}_{R}
+
\sqrt p_{\mathrm{LR}}\ket{1}_{R},
\\
\ket{v_1}^{(p_{\mathrm{LR}})}_{R}
&=
\sqrt p_{\mathrm{LR}}\ket{0}_{R}
+
\sqrt{1-p_{\mathrm{LR}}}\ket{1}_{R}.
\label{eq:v01}
\end{align}
Using Eqs.~\eqref{eq:sqrt-Lambda0-MDI} and
\eqref{eq:sqrt-Lambda1-MDI}, the action of $K^{(p_{\mathrm{LR}})}$ on the
computational basis of
$\mathcal H_{A}\otimes\mathcal H_{B}$ is
\begin{align}
K^{(p_{\mathrm{LR}})}
\ket{xy}_{AB}
=
\begin{cases}
\ket{v_x}^{(p_{\mathrm{LR}})}_{R}\otimes
\ket{xy}_{AB},
&
x,y\in\{0,1\},
\\[1mm]
0,
&
\text{otherwise}.
\end{cases}
\label{eq:K-action-fullbasis}
\end{align}
Consequently,
\begin{align}
\mathrm{Ran}(K^{(p_{\mathrm{LR}})})
=
\mathrm{span}
\left\{
\ket{v_x}^{(p_{\mathrm{LR}})}_{R}\otimes
\ket{xy}_{AB}
:
x,y\in\{0,1\}
\right\}
\equiv
\mathrm{supp}
[
\mathcal G^{(p_{\mathrm{LR}})}(I_{AB})]
\label{eq:supp-G-noisy}
\end{align}
The four vectors in Eq.~\eqref{eq:supp-G-noisy} are mutually
orthonormal, hence,
\begin{align}
\mathrm{rank}
\left[
\mathcal G^{(p_{\mathrm{LR}})}(\mathbb I_{AB})
\right]
=
4.
\end{align}

We introduce two two-dimensional reduced registers
$\bar A$ and $\bar B$ and choose as facial-reduction isometry
\begin{align}
W_{G,p_{\mathrm{LR}}}
:=
\sum_{x,y=0,1}
\left(
\ket{v_x}^{(p_{\mathrm{LR}})}_{R}\otimes
\ket{xy}_{AB}
\right)
\bra{xy}_{\bar A\bar B}.
\label{eq:WG-noisy}
\end{align}
Its four columns are precisely the orthonormal basis of the support
identified in Eq.~\eqref{eq:supp-G-noisy}. Therefore,
\begin{align}
W_{G,p_{\mathrm{LR}}}^\dagger W_{G,p_{\mathrm{LR}}}
&=
\mathbb {I}_{[\bar A\bar B]},
\\
W_{G,p_{\mathrm{LR}}}W_{G,p_{\mathrm{LR}}}^\dagger
&=
\Pi_{
\mathrm{supp}[
\mathcal G^{(p_{\mathrm{LR}})}(\mathbb I_{AB})]
}.
\end{align}
Thus $W_{G,p_{\mathrm{LR}}}$ satisfies
Eqs.~\eqref{eq:FR-isometry-condition} and
\eqref{eq:FR-support-condition} and is a facial-reduction isometry for
$\mathcal G^{(p_{\mathrm{LR}})}$.

\subsubsection{Facial reduction of
$\mathcal Z\circ\mathcal G^{(p_{\mathrm{LR}})}$.}

We now apply the same construction to the pinched map. Using \eqref{eq:Z-noiseless} and \eqref{eq:G-noisy-MDI}
we can write the composed map as
\begin{align}
\left(
\mathcal Z
\circ
\mathcal G^{(p_{\mathrm{LR}})}
\right)(\cdot)
&=
\sum_{r=0,1}
Q_r K^{(p_{\mathrm{LR}})}
(\cdot)
K^{(p_{\mathrm{LR}})\dagger} Q_r
\nonumber\\
&=
\sum_{r=0,1}
K_{ZG,r}^{(p_{\mathrm{LR}})}
(\cdot)
K_{ZG,r}^{(p_{\mathrm{LR}})\dagger},
\label{eq:ZG-Kraus}
\end{align}
where we have defined
\begin{align}
K_{ZG,r}^{(p_{\mathrm{LR}})}
:=
Q_r K^{(p_{\mathrm{LR}})},
\qquad r=0,1.
\label{eq:KZG-def}
\end{align}
Thus, in contrast to $\mathcal G^{(p_{\mathrm{LR}})}$, which has a single Kraus
operator, the pinched map has two Kraus operators. Applying
Eq.~\eqref{eq:FR-support-Kraus}, its support is therefore
\begin{align}
\mathrm{supp}
\left[
\left(
\mathcal Z
\circ
\mathcal G^{(p_{\mathrm{LR}})}
\right)
(\mathbb  I_{AB})
\right]
=
\mathrm{span}_{r=0,1}
\mathrm{Ran}
\left(
K_{ZG,r}^{(p_{\mathrm{LR}})}
\right).
\label{eq:supp-ZG-Kraus}
\end{align}

As for the previous map, these ranges can be determined by evaluating
the Kraus operators on the computational basis of the input space.
Using Eq.~\eqref{eq:K-action-fullbasis}, for
$x,y\in\{0,1\}$ we have
\begin{align}
K_{ZG,r}^{(p_{\mathrm{LR}})}
\ket{xy}_{AB}
&=
Q_r
\left(
\ket{v_x}^{(p_{\mathrm{LR}})}_{R}
\otimes
\ket{xy}_{AB}
\right)
\nonumber\\
&=
\langle r|v_x\rangle^{(p_{\mathrm{LR}})}\,
\ket{r}_{R}
\otimes
\ket{xy}_{AB},
\label{eq:KZG-action}
\end{align}
whereas all basis vectors outside the key-generation subspace are
annihilated.

For $0<p_{\mathrm{LR}}<1/2$, all coefficients
$\langle r|v_x\rangle$ are nonzero. Explicitly,
\begin{align}
\langle 0|v_0\rangle^{(p_{\mathrm{LR}})}
&=
\sqrt{1-p_{\mathrm{LR}}},
&
\langle 1|v_0\rangle^{(p_{\mathrm{LR}})}
&=
\sqrt p_{\mathrm{LR}},
\\
\langle 0|v_1\rangle^{(p_{\mathrm{LR}})}
&=
\sqrt p_{\mathrm{LR}},
&
\langle 1|v_1\rangle^{(p_{\mathrm{LR}})}
&=
\sqrt{1-p_{\mathrm{LR}}}.
\end{align}
Hence, for each fixed value of $r$,
\begin{align}
\mathrm{Ran}
\left(
K_{ZG,r}^{(p_{\mathrm{LR}})}
\right)
=
\mathrm{span}
\left\{
\ket{r}_{R}
\otimes
\ket{xy}_{AB}
:
x,y\in\{0,1\}
\right\}.
\label{eq:Ran-KZG}
\end{align}
Since the two values $r=0,1$ correspond to orthogonal subspaces of
$R$, Eq.~\eqref{eq:supp-ZG-Kraus} gives
\begin{align}
\mathrm{supp}
\left[
\left(
\mathcal Z
\circ
\mathcal G^{(p_{\mathrm{LR}})}
\right)
(I_{AB})
\right]
=
\mathrm{span}
\left\{
\ket{r}_{R}\otimes
\ket{xy}_{AB}
:
r,x,y\in\{0,1\}
\right\}.
\label{eq:supp-ZG-noisy}
\end{align}
The eight vectors in Eq.~\eqref{eq:supp-ZG-noisy} are mutually
orthonormal. Therefore,
\begin{align}
\mathrm{rank}
\left[
\left(
\mathcal Z
\circ
\mathcal G^{(p_{\mathrm{LR}})}
\right)
(\mathbb I_{AB})
\right]
=
8,
\qquad
0<p_{\mathrm{LR}}\le1/2.
\end{align}
Using the reduced registers \(\bar A\) and \(\bar B\), we choose the following facial-reduction isometry:
\begin{align}
W_{ZG,p_{\mathrm{LR}}}
:=
\sum_{r,x,y=0,1}
\left(
\ket{r}_{R}\otimes
\ket{xy}_{AB}
\right)
\left(
\bra{r}_{ R}\otimes
\bra{xy}_{\bar A\bar B}
\right).
\label{eq:WZG-noisy}
\end{align}
Its columns are exactly the eight orthonormal vectors spanning the
support in Eq.~\eqref{eq:supp-ZG-noisy}. Consequently,
\begin{align}
W_{ZG,p_{\mathrm{LR}}}^\dagger W_{ZG,p_{\mathrm{LR}}}
&=
\mathbb I_{R\bar A\bar B},
\\
W_{ZG,p_{\mathrm{LR}}}W_{ZG,p_{\mathrm{LR}}}^\dagger
&=
\Pi_{
\mathrm{supp}
\left[
\left(
\mathcal Z
\circ
\mathcal G^{(p_{\mathrm{LR}})}
\right)
(\mathbb I_{AB})
\right]
}.
\end{align}
Thus $W_{ZG,p_{\mathrm{LR}}}$ satisfies
Eqs.~\eqref{eq:FR-isometry-condition} and
\eqref{eq:FR-support-condition}, and is therefore a
facial-reduction isometry for the pinched map.
The two maps therefore require different facial reductions:
$\mathcal G^{(p_{\mathrm{LR}})}$ has a four-dimensional effective output space,
whereas
$\mathcal Z\circ\mathcal G^{(p_{\mathrm{LR}})}$ has an
eight-dimensional effective output space for $0<p_{\mathrm{LR}}\le1/2$. Finally, the isometry relating the two reduced output spaces is
\begin{align}
\widehat S_{p_{\mathrm{LR}}}
:=
W_{ZG,p_{\mathrm{LR}}}^\dagger W_{G,p_{\mathrm{LR}}}.
\label{eq:S-noisy}
\end{align}
Using Eqs.~\eqref{eq:WG-noisy} and
\eqref{eq:WZG-noisy}, its action on the computational basis of the
four-dimensional reduced space is
\begin{align}
\widehat S_{p_{\mathrm{LR}}}
\ket{xy}_{\bar A\bar B}
=
\ket{v_x}^{(p_{\mathrm{LR}})}_{ R}
\otimes
\ket{xy}_{\bar A\bar B},
\qquad x,y\in\{0,1\},
\end{align}
and therefore
\begin{align}
\widehat S_{p_{\mathrm{LR}}}
=
\sum_{x,y=0,1}
\left(
\ket{v_x}^{(p_{\mathrm{LR}})}_{ R}
\otimes
\ket{xy}_{\bar A\bar B}
\right)
\bra{xy}_{\bar A\bar B}.
\label{eq:S-noisy-explicit}
\end{align}

\begin{remark}

For $p_{\mathrm{LR}}=0$, the pinched map has a four-dimensional support, coinciding with that of the noiseless key map. Therefore, the same facial-reduction isometry can be chosen for both maps, yielding $S_0=\mathbb I_4$. The  eight-dimensional construction presented above therefore applies to $p_{\mathrm{LR}}>0$. This is consistent with the noiseless MDI construction of Ref.~\cite[Appendix D]{pereira2025optimalkeyratesquantum}, where both maps are reduced using a common isometry.
\end{remark}

\end{document}